\documentclass[12pt]{article}
\usepackage{a4wide}
\usepackage{amsmath,amssymb,bbm}
\usepackage{multirow}
\begin{document}
\setcounter{page}{1}
%

%%%%%%%%%%%%%%%%%%%%%%%%%%%%%%%%

%

%%%%%%%%%%%%%%%%%%%%%%%%%%%%%%%%%%%%
%Without pictures use this macro
\def\pct#1{(see Fig. #1.)}
%%%%%%%%%%%%%%%%%%%%%%%%%%%%%%%%%%%
%With pictures use this macro
%\def\pct#1{\input epsf \centerline{ \epsfbox{#1.eps}}}

%%%%%%%%%%%%%%%%%%%%%%%% FRONT PAGE %%%%%%%%%%%%%%%%%%%%%%%%%%%%%%%%%%%%%
\begin{titlepage}
\hbox{\hskip 12cm KCL-MTH-08-01  \hfil}
 \hbox{\hskip 12cm KUL-TF-08/02 \hfil}
%\hbox{\hskip 12cm \hfil}
%\hbox{\hskip 12cm DRAFT $$Revision: 1.61 $$ \hfil}
%\hbox{\hskip 12cm $$Author: t58 $$ \hfil}
%\hbox{\hskip 12cm $$Date: 2003/10/31 14:37:00 $$ \hfil}
%\end{flushright}
\vskip 1.4cm
\begin{center}  {\Large  \bf  Real Forms of  Very Extended Kac-Moody Algebras \\ \vspace{0.4cm}
and Theories with Eight Supersymmetries }

\vspace{2cm}

{\large \large Fabio Riccioni ${}^1$,  \ Antoine Van Proeyen  ${}^2$ \ and \ Peter West ${}^1$}
\vspace{0.7cm}

{\sl ${}^1$ Department of Mathematics,  King's College London  \\
\vspace{0.15cm} Strand \  London \  WC2R 2LS \  UK}
 \vskip .75cm
{\sl ${}^2$ Instituut voor Theoretische Fysica, Katholieke Universiteit
Leuven \\ \vspace{0.15cm} Celestijnenlaan 200D \ B-3001 \ Leuven \
Belgium}

 \vskip 3cm

\end{center}

\abstract{We consider all theories with eight supersymmetries whose reduction to three dimensions gives rise
to scalars that parametrise symmetric manifolds. We conjecture that these theories are non-linear
realisations of very-extended Kac-Moody algebras for suitable choices of real forms. We show for the most
interesting cases that the bosonic sector of the supersymmetric theory is precisely reproduced by the
corresponding non-linear realisation.}

\vfill
\end{titlepage}
%%%%%%%%%%%%%%%%%%%%%%%%%%%%%%%%%%%%%%%%%%%%%%%%%%%%
\makeatletter \@addtoreset{equation}{section} \makeatother
\renewcommand{\theequation}{\thesection.\arabic{equation}}
\addtolength{\baselineskip}{0.3\baselineskip}
%%%%%%%%%%%%%%%%%%%%%%%%%%%%%%%%%%%%%%%%%%%%%%%%%%%%

\section{Introduction}
The original  $E_{11}$ conjecture \cite{1} concerned the eleven dimensional supergravity theory,  the ten
dimensional IIA and IIB supergravity theories \cite{2} and their dimensional reductions. It was later
realised that $E_{11}$ is just one of a class of special algebras called very extended algebras \cite{3}.
Such algebras can be constructed from each finite dimensional semi-simple Lie algebra. If the latter algebra
is denoted by $G$, then the very extended algebra is denoted by $G^{+++}$. It was proposed that the bosonic
string effective action generalised to $D$ dimensions is associated with the non-linear realisation of an
algebra originally denoted by $K_{27}$ in the 26-dimensional case and which was later identified with
$D_{D-2}^{+++}$ \cite{1}, and similarly gravity in $D$ dimensions is associated with $A_{D-3}^{+++}$
\cite{4lambertwest}. Indeed, it was proposed to consider non-linear realisations for all $G^{+++}$ algebras
\cite{5englert} and their low level content was found in \cite{6axel}. Clearly not all these theories possess
supersymmetry,  but the non-linear realisation associated with $E_{11}$, {\it i.e.} $E_8^{+++}$, is at low
levels the bosonic sector of maximal supergravity theories invariant under thirty two supersymmetries
\cite{1}. Also in all the other cases the low level content of the non-linear realisation agrees with that of
the theory to which it is associated with \cite{1,2,4lambertwest,6axel}. In particular, the non-linear
realisation of $G_2^{+++}$ has the bosonic field content of ${\cal N}=2$ supergravity in five dimensions,
that is invariant under eight supersymmetries, while the non-linear realisation of $F_4^{+++}$ is a six
dimensional ${\cal N}=(1,0)$ theory, which again has eight supersymmetries \cite{6axel}.
\par
In all the above cases the  real form of the very extended algebra used to construct the non-linear
realisations is the one which has the maximal number of non-compact generators, called the maximally
non-compact real form. The corresponding local sub-algebra of the non-linear realisation is the maximal
compact sub-algebra which for this real form is the one invariant under the Cartan involution. This is also
the case for the ten dimensional ${\cal N}=1$ supergravity theory which was  found to be associated with the
very extended algebra $D_8^{+++}$, as well as ten-dimensional ${\cal N}=1$ supergravity coupled to one
vector, corresponding to the addition of one Yang-Mills multiplet, which is associated to the very extended
algebra $B_{8}^{+++}$. However, if  one adds $n$, for $n\ge 1$, vector fields then the associated very
extended Kac-Moody algebra is $D_{8+{n\over 2} }^{+++}$ for $n$ even and $B_{8+{(n-1)\over 2}}^{+++}$ for $n$
odd \cite{7igor} and the required real form of these very extended algebras is not maximally non-compact.
\par
Different real forms of a complex algebra have different numbers of compact generators and so different
compact sub-algebras. The local sub-algebra used in the non-linear realisation is the maximal compact
sub-algebra of the real form being considered and the fields in the non-linear realisation are found by
writing down a general group element which depends on space-time and remove parts of it using the local
subgroup. The coefficients of the generators that remain are the fields of the theory. Clearly, if one uses a
different real form, the local sub-algebra is different and as a result so will be the group element in the
non linear realisation and consequently the corresponding field content. Furthermore, the dynamics is just
that which is invariant under rigid transformations of the algebra which is used to define the non-linear
realisation and also under the local sub-algebra. Clearly, a different choice of local sub-algebra also
affects the dynamical equations. Hence, different real forms lead to different non-linear realisations which
are physically inequivalent.
\par
This is apparent in the above examples as the algebra $D^{+++}_{m}$ is associated with the bosonic string in
$m+2$ dimensions and also with ten-dimensional ${\cal N}=1$ supergravity coupled to $2(m-8)$ Yang-Mills
multiplets. The different field content of these theories arises from the fact that the real forms used are
different. Deleting the first three nodes of the Dynkin diagram of $D^{+++}_{m}$ we find the finite
dimensional algebra $D_m$, but while the bosonic string corresponds to the  maximally non-compact real form
$SO(m,m)$ of this algebra with local sub-algebra $SO(m)\otimes SO(m)$, ten-dimensional ${\cal N}=1$
supergravity coupled to $2(m-8)$ Yang-Mills multiplets corresponds to the real form $SO(8,2m-8)$ with local
sub-algebra $SO(8) \otimes SO(2m-8)$ \cite{7igor}.

Different real forms of finite dimensional semi-simple Lie algebras have come up in the context of
supergravity theories in the past. In particular,  certain supergravity theories have scalar fields that
belong to cosets of groups that are not in their maximally non-compact form. The supergravity  theories with
eight supersymmetries that exist in six dimensions and less and the spaces to which their scalars belong have
been the subject of much study. In particular in
\cite{gunaydinsierratownsend,specialkahler,romans,castellanidauriaferrara} the geometry of the spaces that
arise has been studied.

The way the cosets and the corresponding non-split real forms are linked under dimensional reduction or
oxidation have been studied in \cite{gunaydinsierratownsend,specialkahler,toinedewit,cremmerjulialupope,
keurentjes,andrianopolietal}. Different real forms of semi-simple Lie algebras have also occurred in the
context of cosmological billiards \cite{henneauxjulia,billiards,fretoine}.

\par
In the original understanding of the $E_{11}$ and related conjectures an important stepping stone was the
result \cite{8westconformal} that the theories under study, such as eleven dimensional supergravity,  are
non-linear realisations. The Kac-Moody algebra is taken to be the smallest such algebra that contains all the
generators and commutators of the algebra that turn up in the non-linear realisation. If one suspects that
the Kac-Moody algebra associated with a given theory is a very extended algebra then one can adopt a short
cut. The three dimensional theory arising from a very extended Kac-Moody algebra is found by decomposing
$G^{+++}$ into the algebras that result in deleting the affine node of its Dynkin diagram. This is the node
usually labelled three. The resulting algebra is $A_2\otimes G$ where the $A_2$ factor is associated with
three dimensional gravity and the factor $G$ is the internal symmetry group in three dimensions. At lowest
levels the resulting three dimensional theory contains only gravity and scalars which for theories with
supergravity are always found to belong to a coset space or non-linear realisation for the group $G$ and a
subgroup $H$. Hence if one suspects that a theory has a formulation at low levels as a non-linear realisation
of a very extended algebra $G^{+++}$ one can find the group $G$ by dimensionally reducing the theory in
question to three dimensions  and read off the group $G$ from the scalar coset and thus the conjectured very
extended algebra $G^{+++}$ of the higher dimensional theory. The reduction also provides the real form of the
algebra $G^{+++}$ as this is inherited from the real form of the algebra $G$ that turns up in the three
dimensional theory. Indeed the subgroup $H$ of the scalar coset is the maximal compact subgroup of $G$ which
tells us which real form of $G$ arose.
\par
Having chosen the very extended Kac-Moody algebra $G^{+++}$ and its real form one will automatically recover
the correct scalar coset in three dimensions, however the field content in all other dimensions is uniquely
predicted by the non-linear realisation of  $G^{+++}$  and one can test if this agrees with the theory in
question or not.
\par
In this paper we consider the theories with eight supersymmetries whose scalars parame\-trise symmetric
manifolds. These theories have been classified in \cite{toinedewit}. We conjecture that all these theories
have an underlying very-extended Kac-Moody symmetry. We identify this Kac-Moody symmetries and verify that
they lead to precisely the correct field content for some of the theories of most interest.

The paper is organised as follows. In section 2 we give a review of theories with eight supersymmetries in
six, five, four and three dimensions. In section 3 we derive the Kac-Moody algebras $G^{+++}$ associated with
different real forms of various Lie algebras, and conjecture their relation with theories with eight
supersymmetries. In section 4 we show that the bosonic field content of these supersymmetric theories exactly
coincides with the one obtained from the $G^{+++}$ non-linear realisation. Section 5 contains a discussion of
our results. An appendix explains how one obtains the representations of the internal symmetry group for all
$p$-forms in $D$ dimensions starting from a decomposition of the adjoint representation of $G^{+++}$.

\section{Review of theories with eight supersymmetries}
In this paper we are interested in theories with eight supersymmetries and in particular their field content
in order to compare it with the predictions of the very extended Kac-Moody algebras. Such theories exist in
dimensions six and less as is readily apparent if one considers the representations of Clifford algebras.
Indeed in six dimensions a Dirac spinor has 8 complex components, and a Weyl condition can be imposed, so
that the corresponding irreducible spinor has 8 real components. A Majorana condition can be imposed on a
$USp(2)$ doublet of spinors, leading again to 8 real components if the spinors are Weyl. Spinors satisfying
this type of Majorana conditions are often called symplectic Majorana spinors. In five dimensions no
condition can be imposed on a single irreducible spinor, which has 8 real components. As in six dimensions, a
symplectic Majorana condition can be imposed on a doublet of spinors in five dimensions.

In six dimensions theories with eight supersymmetries are minimal because a spinor with eight real components
is irreducible. Minimal supersymmetric theories in six dimensions are usually called ${\cal N} =(1,0)$
theories to denote the fact that the corresponding supercharge is a Weyl spinor and therefore these theories
are chiral. The six-dimensional supergravity multiplet consists of the vierbein $e_\mu{}^a$, a 2-form
$A_{a_1a_2}^{(-)}$ and a gravitino $\psi_{a\alpha i}$. The 2-form has a field strength which is anti-self
dual while the gravitino is a $USp(2)$ symplectic Majorana-Weyl spinor, hence the $i=1,2$ index. In six
dimensions there are also tensor and vector multiplets. The field content of the tensor multiplet is given by
a 2-form $A_{a_1a_2}^{(+)}$ and a scalar $\phi$, together with a symplectic Majorana-Weyl spinor whose
chirality is opposite to the one of the gravitino. The field strength of the 2-form is self-dual and the
scalar $\phi$ is real. The vector multiplet consists of a vector $A_a$ and a symplectic Majorana-Weyl spinor
of the same chirality of the gravitino. There is  also a hyper-multiplet which consists of two fermions and
four scalars, but we will not consider this and its dimensional reductions further in this paper.
\par
If we denote by $n^{(6)}_T$ and $n^{(6)}_V$ the number of tensor and vector multiplets respectively, then the
bosonic content of the six dimensional theory is given by
  $$
  (e_\mu{}^a,A_{a_1a_2}^{(-)}; A_{a_1 a_2}^{(+)} (n^{(6)}_T), \phi (n^{(6)}_T); A_a (n^{(6)}_V)) \eqno(2.1)
  $$
where the numbers $n^{(6)}_T$ and $n^{(6)}_V$ in brackets denote the number of such fields. The scalars
parametrise the coset $SO(n^{(6)}_T,1)\over SO(n^{(6)}_T)$ \cite{romans}.
\par
We now consider the five dimensional case. As in six dimensions, theories with eight supersymmetries are
minimal in five dimensions. Nonetheless, they are usually called ${\cal N}=2$ theories. The dimensional
reduction of the six-dimensional gravity multiplet gives in five dimensions the gravity multiplet together
with one vector multiplet, which has the bosonic field content $(A_a,\varphi)$ where the scalar $\varphi$ is
real, while the dimensional reduction to five dimensions of the six dimensional tensor and vector multiplets
both give rise to the vector multiplet in five dimensions, using the fact that a 2-form is dual to a vector
in five dimensions. The five dimensional supergravity multiplet has the bosonic content $(e_\mu{}^a, B_a)$.
As a result the dimensional reduction of the six dimensional theory to five dimensions has the field content
  $$
  (e_\mu{}^a,B_a , A_a (n^{(5)}_V), \varphi (n^{(5)}_V)) \eqno(2.2)
  $$
where $n^{(5)}_V= n^{(6)}_T+n^{(6)}_V+1$. In the case in which the five-dimensional theory does not have a
six-dimensional origin, still equation (2.2) holds with arbitrary $n^{(5)}_V$.

The complete classification of the ${\cal N}=2$ five-dimensional massless theories describing supergravity
coupled to $n^{(5)}_V$ vector multiplets was achieved in \cite{gunaydinzagermann}. The scalars parametrise
manifolds which are called very special real. For the cases in which the scalar manifold is a symmetric space
a complete classification was derived long time ago in \cite{gunaydinsierratownsend}, and the corresponding
cosets are
  $$
  {SO(n^{(5)}_V-1,1) \over SO(n^{(5)}_V-1)}\times SO(1,1)  \eqno(2.3)
  $$
  $$
  SL(3,R)/ SO(3)\qquad (n^{(5)}_V=5) \eqno(2.4)
  $$
  $$
  SL(3,{C})/ SU(3)\qquad (n^{(5)}_V=8) \eqno(2.5)
  $$
  $$
  SU^{*}(6)/ USp(6)\qquad (n^{(5)}_V=14) \eqno(2.6)
  $$
  $$
  E_{6(-26)}/ F_{4}\qquad \qquad (n^{(5)}_V=26) \eqno(2.7)
  $$
  $$
  {SO(1,n^{(5)}_V ) \over SO( n^{(5)}_V )}  \quad . \eqno(2.8)
  $$

In four dimensions the minimal amount of supersymmetries that a supersymmetric theory can have is four,
corresponding to the fact that an irreducible spinor representation has four real components in four
dimensions. Therefore theories with eight supersymmetries are called ${\cal N}=2$ theories. The gravity
multiplet has a bosonic field content which consists of $(e_\mu{}^a,B_a)$, while the vector multiplet, giving
rise to the ${\cal N}=2$ Yang-Mills theory, has the bosonic field content $(A_a, \phi_1\pm i\phi_2)$, where
$\phi_1$ and $\phi_2$ are real. The dimensional reduction of the above six dimensional theory to four
dimensions leads to ${\cal N}=2$, $D=4$ supergravity coupled to $n^{(4)}_V$  ${\cal N}=2$ Yang-Mills
multiplets where, in case the theory can be obtained from a reduction from 6 dimensions, $n^{(4)}_V=
n^{(6)}_T+n^{(6)}_V+2$. The field content is
  $$
  (e_\mu{}^a,B_a , A_a (n^{(4)}_V), (\phi_1\pm i\phi_2) (n^{(4)}_V)) \quad .  \eqno(2.9)
  $$
Equation (2.9) also holds in the case in which the four-dimensional
theory has no six or five-dimensional origin. The special K{\"a}hler spaces
that are parametrised by the scalars have been widely studied in the
literature \cite{specialkahler,castellanidauriaferrara}. In particular,
in \cite{cremmervanproeyen} the following possible non-compact symmetric
spaces were given:
  $$
  SU(n^{(4)}_V,1) / [ SU(n^{(4)}_V)\times U(1) ]  \eqno(2.10)
  $$
  $$
  SO(n^{(4)}_V-1,2) / [ SO(n^{(4)}_V-1)\times SO(2) ] \otimes SU(1,1)/ U(1)  \eqno(2.11)
  $$
  $$
  [ SU(1,1) / U(1) ]^3 \qquad \quad (n^{(4)}_V  = 3 )\eqno(2.12)
  $$
  $$
  Sp(6,R)/U(3)\qquad \quad (n^{(4)}_V  = 6 ) \eqno(2.13)
  $$
  $$
  SU(3,3)/S[ U(3) \times U(3) ] \qquad \quad (n^{(4)}_V  = 9 ) \eqno(2.14)
  $$
  $$
  SO^*(12)/U(6) \qquad \quad (n^{(4)}_V  = 15 ) \eqno(2.15)
  $$
  $$
  E_{7(-26)} /E_6 \times SO(2) \qquad \quad (n^{(4)}_V  = 27 ) \quad . \eqno(2.16)
  $$
%The massless four dimensional ${\cal N}=2 $ theories describing
%supergravity coupled to $n^{(4)}_V$ vector multiplets has been classified
%in.
In the case in which the theory has a four-dimensional origin, the
manifold that the scalars parametrise is called very special K{\"a}hler
\cite{dewittoine2}.

Finally, we perform the reduction to three dimensions. In three dimensions an irreducible spinor has two real
components, and thus theories with eight supersymmetries are called ${\cal N}=4$. In three dimensions,
gravity is not a propagating degree of freedom, while vectors are dual to scalars. The ${\cal N}=4$
hyper-multiplet in three dimensions consists of four real scalars and four Majorana spinors. The dimensional
reduction of the six dimensional theory to three dimensions gives gravity coupled to the real scalars $\tilde
\phi$, and thus the bosonic field content is
  $$
  (e_\mu{}^a, \tilde \phi (4 n^{(3)}_H)) \eqno(2.17)
  $$
where $n^{(3)}_H$ is the number of hyper-multiplets, and in case of reduction from 6 dimensions, $n^{(3)}_H =
n^{(6)}_T+n^{(6)}_V+3$. The scalars parametrise quaternionic manifolds. If the three-dimensional theory has a
four-dimensional origin, the corresponding manifold is called special quaternionic, while is it also has a
five-dimensional origin it is called very special quaternionic. The quaternionic symmetric spaces are
  $$
  USp(2 n^{(3)}_H,2) / [ USp(2 n^{(3)}_H)\times USp(2) ]  \eqno(2.18)
  $$
  $$
  SU(n^{(3)}_H,2) / [ SU(n^{(3)}_H)\times SU(2) \times U(1) ]   \eqno(2.19)
  $$
  $$
  SO (4,n^{(3)}_H) /[ SO(4) \times SO(n^{(3)}_H) ] \eqno(2.20)
  $$
  $$
  E_{6(2)} / [ SU(6) \times SU(2) ]   \qquad \quad (n^{(3)}_H  = 10 ) \eqno(2.21)
  $$
  $$
  E_{7(-5)}/[SO(12) \times SU(2)  ] \qquad \quad (n^{(3)}_H  = 16 ) \eqno(2.22)
  $$
  $$
  E_{8(-24)} / [E_7 \times SU(2) ] \qquad \quad (n^{(3)}_H  = 28 ) \eqno(2.23)
  $$
  $$
  F_{4(4)} / [ USp(6) \times SU(2) ] \qquad \quad (n^{(3)}_H  = 7 )  \eqno(2.24)
  $$
  $$
  G_{2(2)} / SO(4) \qquad \quad  (n^{(3)}_H  = 2 )       \quad . \eqno(2.25)
  $$
\par
We refer the reader to reference \cite{toine} for a review on the scalar manifolds of theories with eight
supersymmetries. The theories with a homogeneous scalar manifold have been labelled by $L(q,P)$ for $q\geq
-3$ and $P\geq 0$ integers. In the case $q=4m$, for integer $m$, there is an extra possibility which will not
be considered here. The parameters $q$ and $P$ are related to the number of tensor and vector multiplets in
the parent six dimensional theory by $n^{(6)}_T= q+1$ and $n^{(6)}_V=P {\cal D}_{q+1}$. Here ${\cal D}_{q+1}$
is the dimension of the irreducible representation of the Clifford algebra in $q+1$ dimensions with positive
signature, and takes the values ${\cal D}_{q+1}=1 \ {\rm for }\ q=-1,0$, ${\cal D}_{q+1}=2 \ {\rm for }\
q=1$, ${\cal D}_{q+1}=4 \ {\rm for }\ q=2$, ${\cal D}_{q+1}=8 \ {\rm for }\ q=3,4$, ${\cal D}_{q+1}=16 \ {\rm
for }\ q=5,,6,7,8$ and ${\cal D}_{q+8}=16{\cal D}_{q}$ \cite{toinedewit,fretoine}.

\section{Relationship between Kac-Moody algebras and \\ theories with eight supersymmetries}

A central role in our considerations will be played by the various real forms of a given Lie algebra defined
over the complex numbers. The classification of semi-simple Lie algebras was originally  carried out when the
algebras are taken to be over the complex numbers as this is a complete field. Consequently, the end result
of the classification, {\it i.e.} the Dynkin diagram, does not specify a preferred real form of the algebra.
A real form of a Lie algebra is just a choice of generators for which the structure constants are real. Given
such a choice we can then take the algebra to be defined over the real numbers. For example, the $A_1$ Dynkin
diagram, which is a single dot,  corresponds to the complex algebra $SL(2,C)$ which has two real forms; the
compact $SU(2)$ algebra and the non-compact $SL(2,R)$ algebra. The representations of these groups have
different properties and this can lead to very different physical effects depending on which real form is
adopted. The possible third real form $SU(1,1)$ is included as it is isomorphic to  $SL(2,R)$. All real forms
of the finite dimensional semi-simple Lie algebras were found by Cartan in 1914. Some references are
\cite{helgason,araki}.
\par
Any finite dimensional complex semi-simple Lie algebra possesses a unique  real form in which all the
generators are compact. Compact means that the scalar product of the generators, defined by the Killing form,
is negative definite. The negative rather than the usually preferred positive nature is just a result of the
particular choice of constant in the Killing form usually adopted by mathematicians. It is given by taking
the generators $\hat U_\alpha=i(E_\alpha+E_{-\alpha})$, $\hat V_\alpha=(E_\alpha-E_{-\alpha})$ and $\hat
H_a=iH_{a}$, where $\alpha$ is any positive root,  $E_{\alpha}$ the corresponding generator and $H_a$ the
elements of the Cartan sub-algebra. The compact nature of the generators follows in an obvious way from the
fact that the only non-zero scalar product between $E_\alpha$ and $E_{-\alpha}$ is given by $(E_\alpha,
E_{-\alpha})=1$ and $(H_a, H_a)=-(\alpha_a,\alpha_a )<0$. We refer to this compact algebra as ${\cal
{G}}^{cp}$ and to its complexification as ${\cal {G}^{{\bf {C}}}}$.
\par
By considering all involutions of the unique compact real form ${\cal {G}}^{cp}$ one can construct all other
real forms of the complex Lie algebra under consideration.  In particular, the real forms are in one to one
correspondence with all those involutive automorphisms of the compact real algebra. By an involution we mean
a map which is an  automorphism ($\theta (AB)=\theta(A)\theta (B)\  \forall \ A,B \ \in  \cal{G} $) which
squares to one ($\theta^2=1$).
\par
Given an involution  ${\theta}$ we can divide the generators of the compact real form ${\cal {G}}^{cp}$ into
those which possess $+1$ and $-1$ eigenvalues of ${\theta}$. We denote these eigenspaces by
$$
    {\cal{G}} = {\cal{K}} \oplus {\hat {{\cal P}}}
\eqno(3.1)
$$
respectively. Since $\theta $  is an automorphism it preserves the structure of the algebra and as a result
the algebra when written in terms of this split must take  the generic form
$$
    [{\cal{K}},\,{\cal{K}}] \subset {\cal{K}},\quad [{\cal{K}},
    {\hat {\cal P}}] \subset {\hat {\cal P}},
\qquad  [{\hat {\cal P}},\, {\hat {\cal P}}] \subset
    {\cal{K}}.
\eqno(3.2)
$$
From the generators ${\cal P}$ we define new generators ${\cal P}=-i{\hat {\cal P}}$, whereupon the algebra
now takes the generic form
$$
    [{\cal{K}},\,{\cal{K}}] \subset {\cal{K}},\quad [{\cal{K}},\,
     { {\cal P}}] \subset  {\cal{P}},\quad [ {\cal{P}},\,
    {\cal{P}}] \subset (-1) {\cal{K}}.
\eqno(3.3)
$$
Thus we find a new real form of the algebra in  which the generators ${\cal K}$ are compact while the
generators ${\cal P}$ are non-compact. This follows from the fact that all the generators in the original
algebra are compact and so have negative  definite scalar product (the Killing form) and as a result of the
change all the generators $ {\cal P}$ will have positive definite scalar products. Clearly,  the  new real
form has maximal compact sub-algebra ${\cal K}$ and this is just the part of the algebra invariant under
$\theta$.
\par
As each real form corresponds to an involution $\theta$ we can write the corresponding real form as ${\cal
{G}}_\theta$.  For the compact real form the involution is just the identity map $I$ on all generators and so
we may write ${\cal {G}}^{cp}={\cal {G}}_{I}$. The number of compact generators is ${\rm dim} (K)$ and the
number of non-compact generators is ${\rm dim} {\cal G}-{\rm dim} {\cal K}$. The character $\sigma$ of the
real form is the number of non-compact minus the number of compact generators and so $\sigma={\rm dim} {\cal
G}-2{\rm dim} K$.
\par
An important real form can be constructed by considering the involution $\theta_c$ which is a linear operator
that takes $E_{\alpha}\leftrightarrow -E_{-\alpha}$ and $H_a \to -H_a$. Clearly, the generators of the
compact real form transform as ${\hat V}_\alpha\  \to \hat{V}_\alpha$, $\hat{U}_\alpha\  \to -\hat{U}_\alpha$
and ${\hat H}_a\ \to -{\hat H}_a$, where $\hat{V}_\alpha=E_\alpha-E_{-\alpha}$,
$\hat{U}_\alpha=E_\alpha+E_{-\alpha}$ and $\hat{H}_a = H_a$. Using this involution we find a real form with
generators
$$
    V_\alpha=\hat V_\alpha,\
    U_\alpha=-i\hat U_\alpha ,\  H_a=-i \hat H_a \quad .
\eqno(3.4)
$$
The $ V_\alpha$ remain compact generators while $ U_\alpha$ and $ H_a$ become non-compact. By abuse of
notation we are denoting with $H_a$ both the Cartan generators of the complex Lie algebra and the Cartan
generators in this particular real form. The maximal compact sub-algebra is just that invariant under the
Cartan involution. Clearly, the non-compact part of the real form of the algebra found in this way contains
all of the Cartan sub-algebra and it turns out that it has the maximal number of non-compact generators of
all the real forms one can construct. It is therefore called the maximally non-compact real form or split
real form. Using the above notation we can write it as ${\cal G}_{\theta_c}$. For  example, the complex Lie
algebra $D_{n}$ has $SO(2n)$ as its unique compact real form and $SO(n,n)$ as its maximally non-compact real
form. For $E_{8}$ the maximally non-compact form is denoted by  $E_{8 (8)}$ and its maximal compact subgroup
is $SO(16)$. The number in brackets is the character of the real form ($8=248-2.120$) and we will use this
notation for all the real forms of the $E_n$ algebras. Taking different non-trivial involutions we find
different real forms. For example, for $SO(p,q)$ the maximal compact sub-algebra is $SO(p)\otimes SO(q)$
while for the real form of $E_8$ denoted by $E_{8(-24)}$ the maximal compact sub-algebra is $E_7\otimes
SU(2)$.
\par
As the involution $\theta$ is an automorphism it preserves the Killing form and as a result $(\theta
(X),\theta (Y))=(X,Y)=-(X,Y)=0$ if $X\in {\cal K}$ and $Y\in {\cal P}$. It also follows form the above
discussion that $(X,\theta (Y))$ is negative definite. In fact one can define a Cartan involution to be an
involution for which this is true. The Cartan involutions for the split form being called in the past papers
of the authors just the Cartan involution as one has so far mainly dealt with the split case.
\par
As we have discussed the Cartan sub-algebra $H$ of ${\cal {G}}_\theta$ can be split between compact
generators ${\cal K}$ and non-compact generators ${\cal P}$. Let us denote the Cartan sub-algebra elements in
${\cal P}$ by $H_{\cal P}=H\cap {\cal P}$. The real rank $r_\theta$ of ${\cal {G}}_\theta$ is the dimension
of $H_{\cal P}$. Clearly, it takes its maximal value for the split case where it equals the rank of ${\cal
{G}}_\theta$.
\par
Rather than consider the action of the involution on the generators of the algebra it is more convenient to
consider its action on the space of roots $\Delta$ of the algebra. In fact this is equivalent to its action
on the Cartan subalgebra $H$ as the roots belong to the dual space $H^*$, but as we will see it is more
illuminating. We can divide the space of roots into those that have  eigenvalues $\pm 1$; $\Delta_c=\{\alpha;
\theta(\alpha)=\alpha\}$ and $\Delta_{s}= \{\alpha; \theta(\alpha)=-\alpha\}$.  The roots  are associated
with the underlying algebra and so can not in general be divided up into eigenspaces of $\theta$ and so there
are roots  that give neither $\pm 1$ under the action of $\theta$, but a combination of roots.
\par
The action of $\theta$,  and so the determination of the real form of a given Lie algebra, can be encoded on
the so called  Tits-Satake diagram. This is the Dynkin diagram of the complex algebra ${\cal G}^{\bf {C}}$
with some of the nodes coloured black. A node is black if its corresponding simple root is in $\Delta_c$ {\it
i.e.} $\theta (\alpha_a)=\alpha_a$ and the rest of the nodes are white. The map $P_\theta$ defined by
$P_\theta(\alpha)={1\over 2} (\alpha-\theta (\alpha))$ is a projection $P_\theta^2=P_\theta$. Clearly, if
$\beta\in \Delta_{s}$,  then $P_\theta(\beta)=0$. In fact $P_\theta(\alpha)$ is just the projection or
restriction of the root $\alpha$ onto the subspace $H_{\cal P}^*$.
\par
Given a  Tits-Satake diagram we know the action of $\theta$ on the simple roots that are black and one can
deduce its action on all the simple roots using the fact that $\theta^2=1$ and that the action of $\theta$
preserves the Cartan matrix. The latter follows from the fact that $\theta$ is an automorphism, so preserves
the Killing form and that the Cartan matrix is constructed from the scalar product on the roots induced from
the Killing form on the dual space to the roots, that is  the Cartan sub-algebra. Hence from the Tits-Satake
diagram  we find the action of $\theta$ on the roots.
\par
An exception to the black and white dots occurs when two simple roots have the same projection, {\it i.e.}
$P_\theta (\alpha_a)=P_\theta (\alpha_b)$. In this case we can draw an curved arrow between nodes $a$ and
$b$.
\par
Having the action of $\theta$ on the roots we can deduce the action on the generators by taking ${\theta}
(H_a)=H_a$ if the label $a$ corresponds to a black dot. This follows from the action on the corresponding
simple root, indeed the action of $\theta$ on all of $H$ follows from the its action on the dual space of
roots. We also take $\theta (E_\alpha)=c_\alpha E_{\theta(\alpha)}$ where $c_\alpha=\pm 1$. The assignment of
the constants $c_\alpha$ is such that $\theta (E_\alpha)= (E_\alpha)$ if $\alpha\in \Delta_ c$ and $\theta
(E_\alpha)= -(E_\alpha)$ if $\alpha\in \Delta_ s$. The action of the remaining generators must be consistent
with the fact that $\theta$ is an automorphism, acting on $[E_\alpha, E_\beta
]=N_{\alpha,\beta}E_{\alpha+\beta}$ we find that
  $$
  c_{\alpha +\beta} N_{\alpha,\beta}= c_\alpha c_\beta N_{\theta(\alpha ),\theta (\beta )} \quad .\eqno(3.5)
  $$
The black nodes are just the ones whose corresponding Cartan sub-algebra element is compact and obviously the
white nodes are those corresponding to the non-compact elements. Clearly, for the  maximal non-compact real
form all the nodes are white as all the elements of the Cartan sub-algebra are non-compact while for the
compact real form all the nodes are black.
\par
Given a real form of a complex Lie algebra, any element $g$ of the associated group can be expressed as
$g=g_c g_{na} g_r$ where $g_c$ is in the maximal compact sub-group, $g_{na}$ is the maximal commuting
non-compact subalgebra, that is $H_{\cal P}$ and $g_r$ is the group whose Lie algebra consists of the
generators which have  the positive roots with respect to $H_{\cal P}$.  This is the Iwasawa decomposition a
description of which can be found in reference \cite{helgason}. Let us be a little more precise. Given the
generators of the real algebra whose roots  are the  eigenvectors of the full Cartan sub-algebra $H$, we can
consider the restriction of the roots to be just the eigenvalue components corresponding to the Cartan
sub-algebra generators in $H_{\cal P}$. We denote this restricted root space by $\Sigma$. We can think of
$\Sigma$ as the dual of the space $H_{\cal P}$. Given an ordering in $\Sigma$, or equivalently  $H_{\cal P}$,
we can then split the roots in $\Sigma$ into those that are positive denoted $\Sigma^+$ and the rest which
are negative, denoted  $\Sigma^-$. The group $g_r$ is just that generated by the Lie algebra consisting of
all generators whose restricted root is in $\Sigma^+$. The action of $\theta$ sends a restricted root
$\lambda$ to  $-\lambda$.

For the case of a maximally non-compact form of the algebra, all the Cartan generators are non-compact and so
$g_{na}$ is just the Cartan sub-algebra while $g_r$ is generated by all positive root generators of the
complex  algebra, that is all the positive root generators in the usual sense. As such, this is the
decomposition for which $ g_{na} g_r$ is just the Borel sub-group. The important point to note is that it is
only for the maximally non-compact real form that all Cartan subalgebra elements of the original algebra
appear in $g_{na}$ and all simple roots can be generated from multiple commutators of the generators
appearing in $g_r$.
\par
The construction of non-linear realisations based on a given algebra is carried out with respect to a
particular real form of a given algebra and the choice of a local sub-algebra.  The choice of local
subalgebra affects the field content and the way the symmetries are realised. The local subalgebra is for the
cases considered in this paper  always chosen to be the maximal compact subalgebra of a the real form being
used. Clearly, the dimension and the properties of this local sub-algebra change from one real form to
another. For example, for $SO(n,n)$ the maximal compact sub-algebra is $SO(n)\otimes SO(n)$ while for
$SO(p,q)$ it is $SO(p)\otimes SO(q)$. Clearly, even if $p+q=2n, p\not=q$ the dimensions of the two cosets are
different and so is the physics resulting from the two non-linear realisations based on the two algebras. In
particular, only for $SO(n,n)$ do all the Cartan subalgebra generators appear in the coset and so correspond
to fields in the non-linear realisation. So far, all algebras considered in the context of the eleven
dimensional supergravity, IIA and IIB supergravity and their formulations in lower dimensions were the
maximal non-compact form of real algebras and so the group element which appears in the non-linear
realisation can be chosen to be that of the Borel subgroup. This is also the case for $D=10$, ${\cal N}=1$
supergravity theory coupled to no or one vector multiplet, but for more than one vector multiplet coupled to
${\cal N}=1$ supergravity one must use  symmetry algebras that are not the maximally non-compact real form
\cite{7igor}.
\par
Important for the original understanding  that $E_{11}$ is a symmetry of the low energy effective actions of
string theory was the formulation of the corresponding supergravity theories as non-linear realisations of an
algebra $G_{11}$. This latter algebra was not a Kac-Moody algebra, but it was conjectured that the
corresponding theory was associated with the  smallest Kac-Moody algebra  that contained all  the generators,
and their commutation relations, of the algebra $G_{11}$ that arose in the non-linear realisation of the
supergravity theory under study. However, unlike $G_{11}$, the conjectured Kac-Moody algebra contains many
more generators.   As a result the non-linear realisation of the Kac-Moody algebra contains many more fields
than the original non-linear realisation and so only at the lowest levels do the two coincide
\cite{1,8westconformal}. As noted above it was realised that all the Kac-Moody algebras found  by considering
maximal supergravities \cite{1}, effective bosonic string actions \cite{1} and gravity \cite{4lambertwest} in
this way were very extended algebras \cite{5englert}. Given a semi-simple finite dimensional algebra $G$ one
constructs the very extended algebra $G^{+++}$ by adding to the Dynkin diagram of $G$ first the affine node
and then the over extended node, which is connected to the affine node by a single line, and finally the very
extended node, which is connected to the over extended node by a single line \cite{5englert}. Thus $G^{+++}$
has rank three more than $G$.  If a theory is a non-linear realisation of $G^{+++}$, the formulation of the
three dimensional theory is given by carrying out the decomposition of $G^{+++}$ by deleting the affine node
which is usually labeled the node three. This breaks $G^{+++}$ into $G\otimes A_2$.  The second factor is the
algebra  $SL(2)$ and it corresponds to the presence of  gravity in the three-dimensional theory. $G$ is the
internal symmetry of the three dimensional theory. In particular, the field content of the non-linear
realisation is found by carrying out the decomposition of $G^{+++}$ into $G\otimes A_2$ and taking into
account the local sub-algebra as described above. Clearly,  at the lowest level we will find  in addition to
gravity a set of scalars which are a non-linear realisation of $G$ with a local sub-algebra $H$ that is just
the local sub-algebra of the full non-linear realisation which lies in  $G$. In fact the scalars are the only
dynamical degrees of freedom of the three dimensional theory.
\par
Clearly if we suspect that a theory in dimension $D$ has an underlying very extended Kac-Moody algebra we can
reduce it to three space-time dimensions, find the coset the scalars belong to and then deduce the
corresponding very extended Kac-moody algebra for the theory in $D$ dimensions is $G^{+++}$ if the scalars in
three dimensions are a non-linear realisations constructed from the algebra $G$. The Kac-Moody algebra now
determines uniquely all the field content of the theory in $D$ dimensions, it being just a consequence of the
decomposition of $G^{+++}$ into the algebra that results by deleting the node usually called $D$ in the
Dynkin diagram. The first test of the conjectured very extended Kac-Moody symmetry is to see if the field
content it predicts actually agrees with the content of the theory under consideration in the dimension of
interest. If this precise test is not true then the conjecture that the theory has an underlying very
extended Kac-Moody algebra is not true.
\par
The scalars in three dimensions will be a non-linear realisation, or coset, of a particular real form of $G$
and the local sub-algebra $H$ will be the maximal compact sub-algebra. We can then conjecture  a $G^{+++}$
that has the corresponding real form. As explained above the real form of an algebra can be encoded in its
Dynkin diagram by colouring some of the nodes black. This means that the corresponding Dynkin diagram of
$G^{+++}$  will have all white dots except for some black dots in the $G$ part of the Dynkin diagram which
coincide with those found in the internal symmetry $G$ of the three dimensional theory. Thus not only can we
deduce from three dimensions the very extended Kac-Moody algebra, but also its real form.
\par
Let us explain how this works  with some examples.  The theory with eight supersymmetries which in six
dimensions has nine tensor multiplets ($n_T^{(6)}=9$) and sixteen  ($n_V^{(6)}=16$) vector multiplets is
associated with  $L(8,1)$. In three dimensions we find that the scalars belong to the non-linear realisation
$E_{8(-24)}$ where the subscript indicates that this is the real form which has the maximal compact
sub-algebra $E_7\otimes SU(2)$. This is the coset space in equation (2.23). Thus we conjecture that this
theory in six dimensions has an extension such that it is the non-linear realisation of $E_{8(-24)}^{+++}$
with the real form in which nodes labelled 7, 8, 9 and 11 are black as in figure 1.
\par
The six dimensional theory with eight supersymmetries with five tensor multiplets ($n_T^{(6)}=5$) and eight
($n_V^{(6)}=8$) vector multiplets is associated with  $L(4,1)$. In three dimensions we find that the scalars
belong to the non-linear realisation $E_{7(-5)}$ where the subscript indices that this is the real form which
has the maximal compact sub-algebra $SO(12)\otimes SU(2)$. This is the coset space in equation (2.22). Thus
we conjecture that this theory in six dimensions  has an extension such that it is the non-linear realisation
of $E_{7(-5)}^{+++}$ with the real form in which nodes 7, 9 and 10  are black as in figure 2.
\par
Alternatively, the six dimensional theory with three tensor multiplets ($n_T^{(6)}=3$) and four vector
multiplets ($n_V^{(6)}=4$), associated with $L(2,1)$, when reduced to three dimensions has scalars which
belong to the coset constructed from $E_{6(2)}$ which has maximal compact subgroup $SU(6)\otimes SU(2)$. This
is the coset space in equation (2.21). As a result we conjecture that this theory is associated with the
non-linear realisation of $E_{6(2)}^{+++}$. The corresponding Dynkin diagram is shown in figure 3.
\par
Other two examples concern the six dimensional theories with eight supersymmetries with one tensor multiplet
and $P$ vector multiplets, associated with $L(0,P)$, that when reduced to three dimensions have  scalars
parametrising a coset of $SO(P+4,4)$ with local subgroup $SO(P+4)\otimes SO(4)$, as in equation (2.20). For
$P$ even, the conjectured Kac-Moody algebra is $D_{{P\over 2}+4{(4)}}^{+++}$, and the corresponding Dynkin
diagram is shown in figure 4. For  $P$ odd, the conjectured Kac-Moody algebra is $B^{+++}_{{P-1 \over 2} + 4
(4)}$, whose Dynkin diagram is shown in figure 5.
\par
The theories giving rise to the coset spaces in equations (2.24) and (2.25) have already been conjectured in
\cite{6axel} to be associated to the $F_{4(4)}^{+++}$ and $G_{2(2)}^{+++}$ non-linear realisations. These
cases, like the $L(0,0)$ and $L(0,1)$ cases above, are special because the corresponding Lie algebra is
maximally non-compact. The six-dimensional theory corresponding to the $F_{4(4)}^{+++}$ non-linear
realisation has two tensor multiplets and two vector multiplets and is associated with $L(1,1)$. The
$F_{4(4)}^{+++}$ Dynkin diagram is shown in figure 21. The theory corresponding to the $G_{2(2)}^{+++}$
non-linear realisation can not be uplifted to six dimensions, as it is evident from the Dynkin diagram of
figure 22. This theory corresponds to pure supergravity in five dimensions.

The $L(-3,P)$ theory, corresponding to the three-dimensional coset of equation (2.18), has a conjectured
Kac-Moody symmetry $C_{P+2}^{+++}$ whose Dynkin diagram is shown in figure 19. As it is evident from the
diagram, this theory can not be uplifted to any dimension above three. Finally, the $L(-2,P)$ theory,
corresponding to the three-dimensional coset of equation (2.18), has a conjectured Kac-Moody symmetry
$A_{P+3}^{+++}$ whose Dynkin diagram is shown in figure 20. The diagram makes it manifest that the highest
dimension in which this theory can live is four.

In the next section we will analyse the $G^{+++}$ non-linear realisations and show that their field content
exactly agrees with the corresponding supergravity theories. We will consider the cases $L(8,1)$, $L(4,1)$
and $L(0,P)$ ($P$ even) explicitly, corresponding to the $E_{8(-24)}^{+++}$, $E_{7(-5)}^{+++}$ and
$D_{{P\over 2}+4{(4)}}^{+++}$ non-linear realisations respectively, but our results apply to all the other
cases as well.

\section{Field content of real forms of $G^{+++}$}
In this section we will test the conjectured Kac-Moody algebras by computing their low level field content
and seeing if it agrees with the actual field content of the theory it is associated with.

\subsection{$E_{8(-24)}^{+++}$ and  $L(8,1)$}

At first sight it would appear that this conjectured Kac-Moody algebra for $L(8,1)$ must be wrong as the ten
and eleven dimensional maximal supergravities also have $E_8^{+++}$ as their corresponding non-linear
realisation. The former uses the maximally non-compact real form, denoted by $E_{8(8)}^{+++}$,  which has a
Dynkin diagram in which all of its nodes are white, while for the $L(8,1)$ theory we are using the real form
$E_{8(-24)}^{+++}$ as illustrated in the Dynkin diagram of figure 1. As we will see in the following, the
fact that the two real forms are different leads to different field contents for the corresponding non-linear
realisations.
\par
The {\bf six dimensional} theory is obtained by taking the decomposition of $E_{8(-24)}^{+++}$ corresponding
to deleting node six in figure 1 leaving the algebra $D_5\otimes A_5$ as shown in figure 6. The latter factor
is the algebra $SL(6)$ and it leads in the non-linear realisation to six dimensional gravity. The internal
symmetry is the real form $SO(9,1)$ as this corresponds to the positions of the black dots in the $D_5$ part
of the Dynkin diagram. The maximal compact subgroup of  $SO(9,1)$ is  $SO(9)$ and so the scalars in six
dimensions belong to the non-linear realisation of $SO(9,1)$ with local subgroup $SO(9)$.
\par
As discussed above to find the theory, in say $D$ dimensions, arising from the non-linear realisation of a
very extended algebra $G^{+++}$ we must first carry out  the decomposition of $G^{+++}$ into the algebra that
remains after the deletion of an appropriate node in the Dynkin diagram of $G^{+++}$. The resulting set of
generators is independent of which real form we take for $G^{+++}$. The non-linear realisation consists of
group elements $g$  which are subject to  transformation $g \to gh$ where $h$ is a local transformation that
belongs to the compact subalgebra. As noted in section three the general group element can be written as
$g=g_c g_{na} g_r$ and so  the group element can be brought to  the form $g=g_{na} g_r$ using this local
transformation. The parameters that appear in the latter group element are just  the fields of the theory. In
the cases studied in this paper the deleted node, labelled c, is a white node and so its corresponding Cartan
sub-algebra element $H_c$ is in $H_{\cal P}$. As such the restricted roots contain a component that is the
eigenvalue of $H_c$. If we adopt an ordering for  $H_{\cal P}$ such that $H_c$ is the first element then a
restricted root will be positive if it arises from a generator which has  positive level with respect to  the
deleted node, {\it i.e.}  $m_c>0$. It follows that the theory will contain fields corresponding to all
generators that have  positive level with respect to the deleted node. In fact,  this is the same set of
fields that occurs in the split case or indeed for any other real form. We note that this consideration does
not apply for level zero generators.
\par
In view of the last remark, the form fields can be computed using techniques similar to  those of reference
\cite{branes} and in the $E_8^{+++}$ case being studied here, the form fields that  arise in the
$E_{8(-24)}^{+++}$ non-linear realisation are  the same as for the maximal compact real form $E_{8(8)}^{+++}$
and can for example be read off from table 5 of reference \cite{us}.
\par
The 1-forms of $E_{8(-24)}^{+++}$ that arise in six dimensions form the spinor representation of $SO(9,1)$,
{\it i.e.} the ${\bf 16}$.  The 2-forms belong to the 10-dimensional representation of $SO(9,1)$. We recall
that in any $G^{+++}$ non-linear realisation every field appears together with its dual. As in six dimensions
2-forms are dual to 2-forms, in this case the 2-forms in the ${\bf 10}$ of $SO(1,9)$ must satisfy
(anti)self-duality conditions. The rank three forms of $E_{8(-24)}^{+++}$ belong to the $\bf \overline{16}$
representation of $SO(9,1)$. These fields are the duals of the 1-forms. The 4-forms are in the ${\bf 45}$
representation of $SO(9,1)$, that is the adjoint. These are duals to the 9 scalars. The apparent
contradiction arising from having more 4-forms than scalars is resolved by remembering that the dynamics is
invariant under the local sub-algebra which at the lowest level is $SO(9)$. Decomposing the ${\bf 45}$ of
$SO(1,9)$ to the $SO(9)$ sub-algebra leads to ${\bf 9\oplus 36}$. The dynamics will set the field strength of
the ${\bf 36}$ to zero and the remaining ${\bf 9}$ will be dual to the scalars.
\par
Thus we find that in six dimensions the non-linear realisation of $E_{8(-24)}^{+++}$ algebra precisely
predicts
  $$
  h_a{}^b ({\bf 1}),\ A_a ({\bf 16}),\ A_{a_1a_2} ({\bf 10}),\ A_{a_1a_2a_3} ({\bf  \overline{16}}), ,\
  A_{a_1\dots a_4} ({\bf 45}) \eqno(4.1)
  $$
where the numbers in brackets denote the representations of $SO(9,1)$ and we find in addition the nine
scalars mentioned above. In the actual six dimensional theory associated with $L(8,1)$ we have nine tensor
multiplets ($n_T^{(6)}=9$) and sixteen ($n_V^{(6)}=16$) vector multiplets and from equation (2.1) in section
two we can read off the field content. The find that we have precise agreement.  The non-linear realisation
of $E_{8(-24)}^{+++}$ also predicts the number of 5-forms $A_{a_1\dots a_5}$ to be in the ${\bf 144}$
representation of $SO(9,1)$ so predicting the presence of 144 gauged supergravities for this theory and the
number of space-filling 6-forms $A_{a_1\ldots a_6}$ to be in the ${\bf 320 \oplus \overline{126} \oplus 10}$
representation.
\par
To find the {\bf five dimensional} theory predicted by $E_{8(-24)}^{+++}$  we must delete node five in fig. 1
to find the algebras $E_{6(-26)}\otimes SL(5)$ as shown in figure 7. The internal symmetry is therefore
$E_{6(-26)}$ as the distribution of the black dots shows. This real form of $E_6$  has $F_4$ as its maximal
compact subgroup. As a result there are 26 scalars in five dimensions and they belong to the non-linear
realisation of $E_{6(-26)}$ with local subgroup $F_4$. The $SL(5)$ factor leads in the non-linear realisation
to the field $h_a{}^b, a,b =1,2,\dots ,5$ which is five dimensional gravity. The $E_{8(-24)}^{+++}$ algebra
leads to 1-forms and 2-forms that are in the ${\bf 27}$ and $\bf \overline{27}$ representations respectively
of $E_{6(-26)}$. The 2-forms are dual to the 1-forms. There are ${78}$ 3-forms in the adjoint of $E_6$ which
are dual to the 26 scalars once one takes account of the above comments.
\par
Hence in five dimensions the non-linear realisation of $E_{8(-24)}^{+++}$ algebra precisely predicts
  $$
  h_a{}^b ({\bf 1}),\ A_a ({\bf 27}),\ A_{a_1a_2} ({\bf \overline{27}}),\ A_{a_1a_2a_3} ({\bf 78}) \eqno(4.2)
  $$
where the numbers in brackets denote the representations of $E_{6(-26)}$ and we have in addition the 26
scalars mentioned above. This is precisely as required as $n_V^{(5)}=9+16+1=26$ and the actual content of the
five dimensional $L(8,1)$ theory is given in equation (2.2). Finally we also have 4-forms $ A_{a_1\dots a_4}$
which belong to the ${\bf 351}$ of $E_{6(-26)}$ and so we expect this theory to have 351 gauged extensions.
We also have 5-forms that belong to the $\bf \overline{1728}\oplus \overline{27}$.
\par
In {\bf four dimensions} we delete the node four of the  $E_{8(-24)}^{+++}$ Dynkin diagram in figure 1 to
leave $E_{7(-25)}\otimes SL(4)$. This leads to the diagram of figure 8. The real form $E_{7(-25)}$ of $E_7$
that arises is the one which has maximal compact sub-algebra $E_6\otimes U(1)$. As such there are  54 scalars
which belong to the non-linear realisation of $E_{7(-25)}$ with local sub-algebra $E_6\otimes U(1)$. In this
non-linear realisation the 1-forms belong to the ${\bf 56}$ representation of $E_{7(-25)}$. They lead to 28
vector fields together with their magnetic duals. The 2-forms belong to the ${\bf 133}$ representation of
$E_{7(-25)}$, which is the adjoint, and are dual to the 45 scalars in the sense discussed above. Indeed, the
${\bf 133}$ of $E_{7(-25)}$ breaks into the $({\bf 27 \oplus \overline{27}})\oplus ({\bf 78 \oplus 1})$ of
$E_6$. The 2-forms  in the first bracket are dual to the scalars while the fields strengths of the latter
will vanish in the dynamics.
\par
Hence in four  dimensions the non-linear realisation of
$E_{8(-24)}^{+++}$ algebra precisely predicts
  $$
  h_a{}^b ({\bf 1}),\ A_a ({\bf 56}),\ A_{a_1a_2} ({\bf 133}) \eqno(4.3)
  $$
where the numbers in brackets denote the representations of $E_{7(-25)}$ and we have in addition the 45
scalars mentioned above. Examining equation (2.3) we see that this is precisely the correct field content of
the $L(8,1)$ theory in four dimensions. The $E_{7(-25)}$ non-linear realisation also predicts that the
deformations forms $A_{a_1a_2a_3}$ belong to the ${\bf 912}$ representation of $E_{7(-25)}$.
\par
Finally we consider the formulation of the $E_{8(-24)}^{+++}$ theory in {\bf three dimensions}. Deleting node
three in figure 1 we find the algebra $E_{8(-24)}\otimes SL(3)$, as shown in figure 9. The scalars belong to
non-linear realisation of $E_{8(-24)}$ which has the maximal compact sub-algebra $E_7\otimes SU(2)$. This is
the correct coset, but this, unlike all the above results in higher dimensions, was guaranteed by the way in
which the very extended algebra and its real form were guessed.
\par
It is also clear why the $L(8,1)$ theory only exists in six dimensions and less. The gravity line is the
$A_{D-1}$ part of the Dynkin diagram  and, as the name suggests,  it is associated with gravity in the $D$
dimensional theory under study.  To actually lead to gravity it must contain the real form of $A_{D-1}$ that
is $SL(D)$ as this form contains all the Cartan sub-algebra elements as non-compact elements and so, in the
non-linear realisation,  it leads to the diagonal components of the metric. Put another way, if we have some
other real form then some of the diagonal components of the metric will be missing.  As such we can not have
a gravity line that contains a black dot. We also demand that the deleted dot be white in view of the
considerations at the beginning of this section. As such the gravity line,  which must begin from the node
labelled one, and the deleted node must all be white nodes and looking at fig 1 one see that the maximal
dimensions is six. Clearly, this applies to all the real forms of the very extended algebras considered in
this paper and it is amusing to verify the upper dimensions is indeed six for the Dynkin diagram of figs 1-5.

\subsection{$E_{7(-5)}^{+++}$ and  $L(4,1)$}

Let us now turn our attention to the theory associated with $L(4,1)$. This six-dimensional supersymmetric
theory has 5 tensor multiplets and 8 vector multiplets and as argued above should be associated with the very
extended algebra $E_{7(-5)}^{+++}$.

To find  the  {\bf six dimensional} theory predicted by the $E_{7(-5)}^{+++}$ non-linear realisation we must
delete node six as in figure 2. This leads to the Dynkin diagram of figure 10. The internal symmetry algebra
is $SO(5,1)\otimes SU(2)$. We note that $SO(5,1)$ is isomorphic to $SU^*(4)$. The maximal compact subalgebra
of $SO(5,1)$ is $SO(5)$, while $SU(2)$ is compact. Thus we find 5 scalars which belong to the coset $SO(5,1)$
with local sub-algebra $SO(5)$. The non-linear realisation of $E_{7(-5)}^{+++}$ has 1-forms that belong to
the $({\bf 4,2})$ representation of $SO(5,1)\otimes SU(2)$. The 2-forms belong to the $({\bf 6,1})$
representation of $SO(5,1)\otimes SU(2)$ which are all either self dual or anti-self dual. The 3-forms of
$E_{7(-5)}^{+++}$ belong to the $({\bf \overline{4},2})$ representation of $SO(5,1)\otimes SU(2)$ and are the
fields dual to the 1-forms. The 4-forms belong to the $({\bf 15,1})$ and $({\bf 1,3})$ representations. The
former decomposes into the representations $({\bf 5,1})$ and $({\bf 10,1})$ of the local sub-group $SO(5)$.
The $({\bf 5,1})$ fields are dual to the scalars while the fields strengths of the latter as well as  the
$({\bf 1,3})$ are set to zero.
\par
To summarise, the non-linear realisation of $E_{7(-5)}^{+++}$ in six dimensions contains the forms
  $$
  h_a{}^b ({\bf 1}),\ A_a ({\bf 4,2}),\ A_{a_1a_2} ({\bf 6,1}),\ A_{a_1a_2a_3} ({\bf \overline{4},2})
  ,\ A_{a_1\dots a_4} ({\bf 15,1})\oplus
  ({\bf 1,3}) \eqno(4.4)
  $$
where the numbers in brackets denote the representations of $SO(5,1)\otimes SU(2)$ and we have in addition
the five scalars mentioned above. This is precisely the content of the $L(4,1)$ as given in equation (2.1)
with 5 tensor multiplets and 8 vector multiplets.

Finally we also predict that the  5-forms  $A_{a_1\dots a_5}$ belong to the ${\bf (4,2)\oplus
(\overline{20},2)}$ representation of $SO(5,1)\otimes SU(2)$. The corresponding field strengths are dual to
mass deformations, and so we expect the same number of gauged supergravities. The space-filling 6-forms
belong to the ${\bf (64,1) \oplus (\overline{10},3) \oplus (6,3) \oplus (6,1) \oplus (6,1)}$ representation.
\par
To find the field content of $E_{7(-5)}^{+++}$  in {\bf five dimensions} we must delete node five in figure
2, as shown in figure 11. The resulting algebra is $SL(5)\otimes SU^*(6)$ where the maximal compact subgroup
of the latter factor is $USp(6)$. The 14 scalars belong to the corresponding coset. The forms fields in the
non-linear realisation of $E_{7(-5)}^{+++}$ appropriate to five dimensions are
  $$
  A_a ({\bf 15}), \ A_{a_1a_2} ({\bf \overline{15}}), \  A_{a_1a_2 a_3} ({\bf 35})\quad .
  \eqno(4.5)$$
The numbers in brackets denote the $SU^*(6)$ representations the form
belongs to. This is in precise agreement with the field content of the
actual $L(4,1)$ theory in five dimensions as can be seen by noticing that
$n_V^{(5)}=n_T^{(6)}+n_V^{(6)}+1=14$ and examining equation (2.2). The
non-linear realisation also predicts 4-forms in the ${\bf 105\oplus 21}$
and so we expect this number of gauged supergravities.  The space-filling
forms predicted by the non-linear realisation belong to the ${\bf 384
\oplus \overline{105} \oplus \overline{15}}$.
\par
The Dynkin diagram of the $E_{7(-5)}^{+++}$ non-linear realisation appropriate to {\bf four dimensions} in
shown in figure 12 where node four of figure 2 has been deleted. The remaining algebra is $SL(4)\otimes
SO^*(12)$. The 30 scalars belong to the coset of $SO^*(12)$ with local subgroup $SU(6)\otimes U(1)$. The form
fields of this four dimensional non-linear realisation are
  $$
  A_a ({\bf 32}), \ A_{a_1a_2} ({\bf 66}), \  A_{a_1a_2 a_3} ({\bf 352}), \  A_{a_1\dots  a_4}  ({\bf
  2079\oplus 462\oplus 66}) \quad .\eqno(4.6)
  $$
The numbers in  brackets denote the $SO^*(12)$ representations. The
1-forms do account for the 16 vectors and their duals, while the 66
2-forms decompose into $SU(6)$ representations as ${\bf 66\to 1\oplus
35\oplus 15\oplus 15}$. The ${\bf 15\oplus 15}$ are dual to the scalars
and the field strengths of the remaining fields are set to zero. We
expect $352$ gauged supergravities, as they are associated to the number
of 3-forms predicted in the non-linear realisation.

Finally, we consider the {\bf three dimensional} case. The Dynkin diagram of the three-dimensional
$E_{7(-5)}^{+++}$ non-linear realisation is obtained deleting node three in figure 2 and leads to the diagram
of figure 13. The remaining algebra is $SL(3) \otimes E_{7(-5)}$, and the maximal compact subgroup of the
latter is $SO(12)\otimes SU(2)$. There are 64 scalars describing the non-linear realisation of $E_{7(-5)}$
with local subgroup $SO(12)\otimes SU(2)$. One can compute the field content in this case, finding precise
agreement with the field content of the three dimensional $L(4,1)$ supersymmetric theory. In particular, the
1-forms belong to the ${\bf 133}$ that is the adjoint of $E_7$, and are related to the scalars by duality.

\subsection{$D_{{P\over 2}+4{(4)}}^{+++}$ and  $L(0,P)$, $P$ even}
We now consider the $L(0,P)$ theory with $P$ even, that was conjectured in section 3 to correspond to the
$D_{{P\over 2}+4{(4)}}^{+++}$ non-linear realisation. We refer to the appendix for a proper explanation of
the computations carried out in this subsection.

Deleting node six of the $D_{{P\over 2}+4{(4)}}^{+++}$ Dynkin diagram in figure 4 we find the diagram of
figure 14. The resulting algebra is $ SO(6,6) \otimes SO(P)$. What is different to the above cases is that
the deletion does not lead to an $SL(D)$ algebra for the space-time part of the remaining algebra, but rather
$SO(6,6)$. As a result we must carry out a further decomposition of this group to $SL(6)$ to find the usual
representations belonging to space-time, the adjoint representation of $SL(6)$ being  associated with the
gravity  line of nodes one to five. The decomposition of $D_{{P\over 2}+4{(4)}}^{+++}$ into representations
of   $SO(6,6)\otimes SO(P)$ is graded by the level $m_c$ associated with node six. This is the number of
times the simple root $\alpha_6$ occurs in the root being considered in the decomposition. At level zero
$m_c=0$ we just have the adjoint representation of $SO(6,6)\otimes SO(P)$, that is $({\bf 66, 1})$ and $({\bf
1, {P(P-1)\over 2}})$ as well as $({\bf 1,1})$. The resulting fields are subject to transformations of the
local subgroup which is $SO(6)\otimes SO(6) \otimes SO(P)$.  Clearly, any scalar fields that might arise in
$({\bf 1, {P(P-1)\over 2}})$, {\it i.e.} the adjoint of $SO(P)$  are  completely removed by the last part of
the local subgroup as $SO(P)$ is compact and thus coincides with its maximal compact subgroup. The adjoint
representation of $SO(6,6)$, {\it i.e} the $({\bf 66, 1})$ breaks into $SL(6)$ representations as ${\bf
66}\to {\bf 35\oplus 15\oplus \overline{15}\oplus 1}$. These correspond in the non-linear realisation to the
graviton $h_a{}^b$, that is ${\bf 35\oplus 1}$, a rank two anti-symmetric tensor $A_{a_1a_2}$, that is $\bf
{15}$, a scalar $\phi$, that is $({\bf 1 ,1})$, while the local sub-algebra removes the ${\bf \overline{15}}$
and the anti-symmetric part of $h_a{}^b$. In fact the $\phi$ can be thought of as belonging to the coset
$SO(1,1)$ with trivial local subgroup.
\par
The fields at the next levels can be found using the decomposition techniques of reference \cite{branes}. In
fact, at the next level, $m_c=1$, there is always an obvious solution for any such reduction of a very
extended algebra. If the resulting algebra after the deletion of node $D$ is $G_1\otimes G_2$ and the deleted
node attaches to the node labelled $i$ of the Dynkin diagram of $G_1$ and the node labelled $j$ of the Dynkin
diagram of $G_2$, then one finds the representation with highest weight $\mu_i\otimes \lambda_j$ where
$\mu_i$ is the fundamental weight of $G_1$ associated with node $i$  and similarly for $\lambda_j$. In all
the cases above $G_1=SL(D)$, the deleted node attaches to the node labelled  $D-1$ and so $i=D-1$ and so one
finds the vector fields. As a result, in the cases above, one finds that the vector fields belong to the
representation of $G_2$ with the highest fundamental weight which is associated with the first node in $G_2$.
The reader can verify that this is indeed the representation for the vectors in all the above cases.
\par
In the case under study in this sub-section  we find that the level one representation that arises in the
decomposition is ${\bf \mu_5\otimes \lambda_7}$. That is the 32 dimensional spinor representation of
$SO(6,6)$ which is valued as a $P$ vector of $SO(P)$. The 32 dimensional spinor representation in question
decomposes into $SL(6)$ representations as ${\bf 32} \to{\bf \overline{6} \oplus 20 \oplus 6}$.  The latter
we recognise as leading to the fields $A_{a_1}$, $A_{a_1a_2a_3}$ and $A_{a_1a_2a_3a_4a_5}$  respectively all
in the vector representation of $SO(P)$. The $P$ 3-forms are the duals of the vectors which are also in the
vector representation. These are the only representations at level one.
\par
At the next level, $m_c=2$, we find the following representations of $SO(6,6)\otimes SO(P)$;
  $$
  {\bf (\mu_4,1)\oplus(1,2\lambda_1)\oplus(\mu_2,\lambda_8)\oplus(1,1) }\eqno(4.7)
  $$
which are labelled in terms of their fundamental weights. We note that $\mu_4$ and $\mu_2$ are  the $495$ and
$66$ dimensional representation of $SO(6,6)$ respectively. To find the fields with space-time indices of
$SL(6)$ we must decompose the representations of $SO(6,6)$ into those of $SL(6)$. This can be achieved by
also deleting node ${P \over 2} +7$ in the $SO(6,6)$ Dynkin diagram. It is straightforward to see that having
done this the number of space-time indices on the generators, or fields, for a representations with highest
weight $\mu=\sum_iq_i\mu_i$ of $SO(6,6)$   is given by $\sum_j (6-j)q_j+6m=m_c+2l$ where $l$ is the number of
times the root of the node labeled ${P \over 2} +7$ occurs in the way the highest weight state occurs and $m$
is the number of blocks of six indices. Among this set of states we are searching for  forms, which excludes
the possibility of having blocks of six indices as well as other indices. If we focus on 2-form fields
$A_{a_1a_2}$ then as $m_c=2$ we must have $l=0$ and so we are really only interested in the $A_5$ part of
$SO(6,6)$. As such we have only one 2-form which occurs at the beginning of the $495$ multiplet. The apparent
scalars in the ${\bf (1,2\lambda_1)}$ actually have a block of six indices and so are not forms in the
required sense.
\par
Hence up to level $m_c\le 2$ and for form fields of rank two or less we have the  fields
  $$
  h_a{}^b({\bf 1}),\ A_{a_1a_2}^{(-)}({\bf 1});\ A_{a_1a_2}^{(+)}({\bf 1}),\ \phi({\bf 1});\
  A_{a_1}({\bf P}),\ A_{a_1a_2a_3}({\bf P})
  \eqno(4.8)
  $$
where the numbers in brackets denote the representations of $SO(P)$. In fact as $m_c\ge 3$ we can not get
more rank three or less forms at higher levels. These are the fields of six dimensional supergravity plus one
tensor multiplet and $P$ vector multiplets. This is indeed what we expect from the bosonic field content of
the $L(0,P)$ theory.
\par
In fact deleting node six is not the only way in which one can get a six dimensional theory. One can also
delete node five and then take the gravity line to  contain nodes 1,2,3,4 and ${P \over 2} +7$ in figure 4.
This leads to the Dynkin diagram in figure 15. The remaining algebra is then $SL(6)\otimes SO(1,P+1)$. It is
straightforward to verify that at level zero, $m_c=0$, one finds $({\bf 1,{(P+2)(P+1)\over 2}})$, $({\bf
35,1} )$ and $({\bf 1,1})$. The last two correspond to gravity and the first to the $P+1$ scalars which
belong to the coset $SO(1,P+1)$ with local sub-group $SO(P)$. At level one, $m_c=1$, we find the
representation $({\bf 15,P+2})$ which corresponds to 2-forms $A_{a_1a_2}$ in the vector representation of
$SO(1,P+1)$. At level two, $m_c=1$, we find the representations $({\bf 15,{(P+2)(P+1)\over 2}})$ and $({\bf
105,1})$. Consequently, in this model we find gravity, $P+2$ self or anti-self dual rank 2-forms and $P+1$
scalars. As a result, the field content it corresponds to is supergravity coupled to $P+1$ tensor multiplets.
\par
Upon dimensional reduction to five dimensions the theories corresponding to the Dynkin diagrams of figure 14
and 15 give the same five dimensional theory as it arises from deleting node five as is evident from the
Dynkin diagram of figure 16. This is very similar to the situation of the well known IIA and IIB theories in
ten dimensions and the way they fit into $E_8^{+++}$ \cite{2} giving rise to a unique nine dimensional theory
upon dimensional reduction. The four-dimensional theory corresponds to deleting node four and gives the
Dynkin diagram of figure 17, and finally the three-dimensional theory arises from deleting node three and
corresponds to the Dynkin diagram of figure 18. It is straightforward to also confirm that by deleting nodes
five, four and three we also recover the correct field content of the bosonic sector of the $L(0,P)$ theories
in five, four and three dimensions respectively.

\section{Discussion}
In this paper we have given substantial evidence to the conjecture that all the theories with eight
supersymmetries that upon reduction to three dimensions give rise to scalars that parametrise symmetric
manifolds have an underlying very extended Kac-Moody symmetry. In particular the bosonic sector of any of
these theories can be derived from the non-linear realisation. We have worked out in detail the $L(8,1)$,
$L(4,1)$ and $L(0,P)$ ($P$ even) cases, and we have found that the bosonic field content of these
supersymmetric theories is precisely reproduced by the non-linear realisations.

Crucial to our analysis are different real forms of very-extended Kac-Moody algebras. We explain the real
forms of the very extended Kac-Moody algebras that we conjecture to describe the various theories with eight
supersymmetries. These are presented using Tits-Satake diagrams. Given this diagram we derive the generators
of the very-extended algebra and the field content of the corresponding non-linear realisation at low levels.

The analysis of the field content of the very extended algebras can also be done for the other cases whose
symmetries we have conjectured. In particular the other theories that live in six dimensions and that give
rise to three-dimensional theories whose scalars parametrise symmetric spaces are $L(0,P)$ for $P$ odd,
corresponding to the $B_{{P-1\over 2} +4}^{+++}$ non-linear realisation whose Dynkin diagram is shown in
figure 5, $L(2,1)$, corresponding to the $E_{6(2)}^{+++}$ non-linear realisation whose Dynkin diagram is
shown in figure 3, and $L(1,1)$, corresponding to the $F_{4(4)}^{+++}$ non-linear realisation whose Dynkin
diagram is shown in figure 21. One can also consider the theories associated to symmetric spaces in three
dimensions that can not be uplifted to six dimensions, namely $L(-3,P)$, corresponding to the $C_{P+2}^{+++}$
non-linear realisation whose Dynkin diagram is shown in figure 19, $L(-2,P)$, corresponding to the
$A_{P+3}^{+++}$ non-linear realisation whose Dynkin diagram is shown in figure 20 (in particular ${\cal N}=2$
4-dimensional supergravity without matter corresponds to the case $P=-1$ of $L(-2,P)$), and minimal
five-dimensional supergravity, corresponding to the $G_{2(2)}^{+++}$ non-linear realisation whose Dynkin
diagram is shown in figure 22. The low-level fields associated to all these Kac-Moody symmetries have been
derived in \cite{6axel}, where it was also shown that the $G_{2(2)}^{+++}$ non-linear realisation describes
the bosonic sector of minimal five-dimensional supergravity and the $F_{4(4)}^{+++}$ non-linear realisation
describes the bosonic sector of the $L(1,1)$ theory.

In \cite{dualE11} it was shown that amongst the infinitely many fields in the non-linear realisation of
$E_{8}^{+++}$, there is an infinite preferred set that describes all possible dualisations of the on-shell
degrees of freedom of the eleven-dimensional supergravity theory. This lifts the infinite set of dualities
that occur in two dimensions to eleven dimensions. All the infinitely many remaining fields in eleven
dimensions have at least one set of ten or eleven antisymmetric indices, and therefore they do not correspond
to on-shell propagating degrees of freedom. This is actually true for any Kac-Moody algebra \cite{duncan},
and thus it applies to all the cases discussed in this paper as well. Therefore although the $G^{+++}$
non-linear realisations discussed in this paper have different real form to those considered before, it is
still the case that all the propagating degrees of freedom of these theories are the infinitely many dual
descriptions of the propagating fields of the corresponding supergravity.

Supersymmetric theories with eight supersymmetries contain exotic supersymmetry representations, like for
instance tensor multiplets in five dimensions. Although 2-forms can be dualised to vectors in five dimensions
in the absence of a potential, for theories with non-trivial vacua this is no longer true. These multiplets
are thus relevant in the context of gauged supergravities. The fact that the $G^{+++}$ non-linear realisation
describes democratically all the fields and the corresponding duals means that it automatically encodes
either description.

In the democratic formulation that arises in the $G^{+++}$ non-linear realisations, turning on a mass
deformation corresponds to having a $D-1$ form whose field strength is dual to the mass and thus is
non-vanishing. In \cite{us,ericembedding} it was shown that all the massive deformations of gauged maximal
supergravities are encoded in the $E_8^{+++}$ non-linear realisation. A similar analysis was carried out in
\cite{eric16} for the case of theories with 16 supersymmetries, corresponding to the $B_m^{+++}$ and
$D_m^{+++}$ non-linear realisations of \cite{7igor}. If this is true also in the case of theories with eight
supersymmetries, this would mean that the number of $D-1$ forms in $D$ dimensions would give in all cases the
number of massive deformations of the supersymmetric theory. Moreover, given that the representation to which
each form belongs does not depend on the particular real form of $G$ being used, this would mean that for
instance the $L(8,1)$ theory associated to $E_{8(-24)}^{+++}$ would possess the same massive deformations in
a given dimension as the $E_{8(8)}^{+++}$ theory, that is associated to the maximally supersymmetric
theories. It would be interesting to investigate in this direction and in particular examine if the fact that
the local subalgebras are different for different real forms affects this result.

The fact that different real forms can be accounted for in the $G^{+++}$ non-linear realisation also leads to
the conjecture that any supergravity theory with more than 16 supersymmetries can be described as a
non-linear realisation for a suitable real form of a very extended $G^{+++}$. In particular, the scalars of
the three-dimensional theory with 18 supersymmetries parametrise the coset $F_{4(-20)} / SO(9)$, and we
conjecture that it is associated to the $F_{4(-20)}^{+++}$ non-linear realisation whose Dynkin diagram is
shown in figure 23. As it is evident from the diagram, this theory only lives in three dimensions. The
supergravity theory with 20 supersymmetries is associated to the $E_{6(-14)}^{+++}$ non-linear realisation,
whose corresponding Dynkin diagram is shown in figure 24. The highest dimension in which this theory exists
is 4, as can be read from the diagram. Finally, the supergravity theory with 24 supersymmetries corresponds
to the $E_{7(-5)}^{+++}$ non-linear realisation whose Dynkin diagram is shown in figure 2. This last example
in particular shows that this real form can lead to separate theories that only differ in the fermionic
sector, so that there should be two different ways of embedding the fermions in the $E_{7(-5)}^{+++}$
non-linear realisation, one giving the $L(4,1)$ theory considered in this paper and one giving the
supergravity theory with 24 supersymmetries. Just like the $L(4,1)$ theory, supergravity with 24
supersymmetry exists in six dimensions and below. The six-dimensional theory is called ${\cal N} =(2,1)$, and
it was originally conjectured in \cite{julia} and later constructed in \cite{dauriaferrarakounnas}. The
reader can check that the bosonic field content of this theory coincides with the one of the
$E_{7(-5)}^{+++}$ non-linear realisation derived in section 4.2. The bosonic string effective action
generalised to $D$ dimensions is associated with the non-linear realisation of the maximally non-compact form
of $D_{D-2}^{+++}$ \cite{1}, and gravity in $D$ dimensions is associated with the maximally non-compact form
of $A_{D-3}^{+++}$ \cite{4lambertwest}. These are examples of real forms which lead to theories that are not
supersymmetric. More generally it is possible that particular real forms of very-extended Kac-Moody algebras
lead to theories with less than eight supersymmetries, or indeed no supersymmetry at all.

\vskip 3cm

\section*{Acknowledgments}
We thank Duncan Steele for discussions. P.W. would like to thank the physics department in Leuven for their
hospitality where this work was begun in 2001. The work of F.R. and P.W. is supported by a PPARC rolling
grant PP/C5071745/1 and the EU Marie Curie, research training network grant HPRN-CT-2000-00122. We further
thank the Galileo Galilei Institute for Theoretical Physics for hospitality and the INFN for partial support.
The work of A.V.P. is supported in part by the European Community's Human Potential Programme under contract
MRTN-CT-2004-005104 `Constituents, fundamental forces and symmetries of the universe', in part by the FWO -
Vlaanderen, project G.0235.05 an in part by the Federal Office for Scientific, Technical and Cultural Affairs
through the `Interuniversity Attraction Poles Programme -- Belgian Science Policy' P6/11-P.

\vskip 2cm

\begin{appendix}
\section{The Calculation of the Form Fields}
In this paper  we have required the decomposition of the adjoint representation of certain Kac-Moody algebras
$G^{+++}$ in terms of representations of $A_{D-1}$ for suitable choices of $D$, were $A_{D-1}$ is a
subalgebra of $G^{+++}$. In this appendix we will show how to algebraically calculate the representations of
the generators with completely antisymmetrised $A_{D-1}$ indices arising in $G^{+++}$. In the non-linear
realisation of $G^{+++}$ these generators are associated to fields with the same $A_{D-1}$ index structure.
This work is carried out by the authors of this paper in collaboration with Duncan Steele.

For no indefinite Kac-Moody algebra is a complete listing of the generators known. However, there is a class
of such algebras called Lorentzian algebras, which includes very extended algebras, whose Dynkin diagram
contains at least one node whose deletion leads to Dynkin diagrams that are those for finite algebras with
possibly one affine algebra, which are more amenable to analysis. Indeed, one can analyse the content of such
Lorentzian algebras in terms of these remaining algebras \cite{3}. Given a very extended algebra $G^{+++}$
the field content of the non-linear realisation it leads to in a given dimension is found by deleting a
particular node and decomposing the adjoint representation of $G^{+++}$ in terms of the representations of
the remaining algebra $G^{+++}_{Del}$, corresponding the the remaining Dynkin diagram after the deletion.
\par
In this appendix we will restrict our attention to the cases where the deletion of  the node in the Dynkin
diagram of $G^{+++}$ corresponds to the decomposition of $G^{+++}$ required into representations of
$G^{+++}_{Del}=G_1 \otimes G_2$ where $G_1$ and $ G_2$ are finite dimensional semi-simple Lie algebras. We
will also restrict our attention to the case of simply laced algebras. The discussion is the same as that
given in reference \cite{branes}, but the emphasis there was on the content of the $l_1$ multiplet, that is
the brane charges, while here we want to focus on the adjoint representation. In essence on takes $m_*=0$ in
that paper. We will eventually consider in detail the case in which $G_1$ is $A_{D-1}$.  However, our methods
are quite general and apply to any semi-simple algebra $G_1$, although when $G_1$ is not  $A_{D-1}$ one must
carry out a further decomposition to this latter algebra to find the field content in terms of familiar
representations. Nonetheless, the analysis  carried out in this appendix is completely general and applies to
any $G_1$ and $G_2$ that can arise in the decomposition of $G^{+++}$.
\par
Let us label the deleted node by $c$.  The simple roots $\alpha_a$ of $G^{+++}$ can be taken to be the simple
roots $\beta_i$ of $G_1$, the simple roots of $\alpha_i$ of $G_2$  and the simple root $\alpha_c$
corresponding to the deleted node $c$. The latter simple root can be written as
  $$ \alpha_c = x - \nu
  \eqno(A.1)
  $$
where $x$ is a vector orthogonal to the root space of $G_1 \otimes G_2$ and
  $$
  \nu =- \sum_i A_{cj} \mu_j  - \sum_i A_{ci} \lambda_i \ \ . \eqno(A.2)
  $$
Here $A_{ab}$ is the Cartan matrix of $G^{+++}$, $\mu_i$ and $\lambda_j$ the fundamental weights of $G_1$ and
$G_2$ respectively and the  nodes of these two algebras are labeled by the same indices $i,j,\ldots$ for
simplicity, although the ranges are different. The vector $\nu$ may be split into $\nu_1$ and $\nu_2$ which
are the parts of $\nu$  belonging to the weight spaces of $G_1$ and $G_2$ respectively. The value of $x^2$ is
determined by the requirement that $\alpha_c^2=2=x^2+\nu^2$.
\par
Using the above expressions, we may write a general root $\alpha$  of $G^{+++}$ as
  $$
  \alpha = m_c\alpha_c + \sum_jn_j\beta_j + \sum_im_i\alpha_i   = m_cx -
\Lambda_{G_1} - \Lambda_{G_2} \eqno(A.3)
  $$
where $m_i$,  $n_i$ and $m_c$ are positive or negative  integers depending if the root $\alpha$ is a positive
or negative root. Also we define the above quantities as
  $$
  \Lambda_{G_1} = m_c\nu_1 - \sum_jn_j\beta_j ,\ \ \ \ \ \ \Lambda_{G_2}
= m_c\nu_2 - \sum_im_i\alpha_i \
  \ .
  \eqno(A.4)
  $$
We note that these two vectors belong to the weight spaces of $G_1$ and $G_2$ respectively.
\par
We will call the integer $m_c$ the level and we will classify  the result of the decomposition into
representations of $G_1 \otimes G_2$ by the level. The level is just the number of times the root $\alpha_c$
occurs in a particular root $\alpha$ being considered. All generators in the algebra $G^{+++}$ can be
constructed from the multiple commutators of the Chevalley generators. As a result, $m_c$
 is just the number
of times the Chevalley generator corresponding to node $c$ occurs in the multiple commutator which leads to
the root of interest.
\par
If a representation of $G_1$ with highest weight $\sum_i q_i\mu_i$, where $q_i$ are the Dynkin indices that
must be positive integers or zero, occurs then this highest weight must occur as one of the possible
$\Lambda_{G_1}$'s that appear as the roots of the $G^{+++}$ vary.  As such, a necessary condition for the
adjoint representation of $G^{+++}$ to contain the highest weight representation of $G_1$ with Dynkin indices
$q_j$ is that
  $$
  \sum_i q_i \mu_i = m_c \nu_1 - \sum_i n_i \beta_i \eqno(A.5)
  $$
where $n_i$ denote the coefficients of the simple roots of the $G_1$ algebra. Taking the scalar product of
both sides of this equation  with $\mu_j$ implies that \cite{branes}
  $$ \sum_i q_i (\mu_i,\mu_j) - m_c(\nu_1,\mu_j) = - n_j \ \ .
  \eqno(A.6)
  $$
In these equations $q_i$, $m_c$ and $n_i$ are positive integers and this places a bound on the possible
highest weights, or Dynkin indices $q_i$ that can occur.
\par
Repeating this procedure for the  $G_2$ algebra, and using $p_i$ to denote the Dynkin indices, we find that
the representation of $G_2$ with highest weight $\sum_ip_i\lambda_i$ occurs provided
  $$
  \sum_i p_i \lambda_i = m_c \nu_2 - \sum_i m_i \alpha_i \ \ . \eqno(A.7)
  $$
Taking the scalar product with $\lambda_k$ we find that \cite{branes}
  $$
  m_k = m_c(\nu_2,\lambda_k) - \sum_i p_i (\lambda_i,\lambda_k) \ \ . \eqno(A.8)
  $$
We note that the occurrence of the highest weights in $G_1$ and $G_2$ is correlated as equations (A.6) and
(A.8) contain the same level $m_c$.
\par
Squaring equation (A.3) gives \cite{branes}
$$
\alpha^2 =  x^2 m_c^2 + \sum_{i,j}q_iq_j (\mu_i,\mu_j) + \sum_{i,j}p_ip_j(\lambda_i,\lambda_j) \ \
.\eqno(A.9)$$ Since $\alpha^2 $ can only take the values $2,0,-2,\ldots$ this again places a constraint on
the allowed representations.
\par
Our task is to analyse equations (A.6), (A.8) and (A.9) to find the possible representations of $G_1 \otimes
G_2$ that can occur in the decomposition of the adjoint representation of $G^{+++}$. Not every solution will
correspond to a root in $G^{+++}$ as these conditions are not as strong as the construction of the algebra
$G^{+++}$ using its definition, that is the multiple commutator of the Chevalley generators subject to the
Serre relations. However, almost all solutions are in fact present in $G^{+++}$ although one does not
discover the multiplicity of the representations using these equations. In the above we have glossed over
some subtle points that are described in more detail in \cite{A8,branes}. The analysis of Lorentzian algebras
in terms of algebras that occur after the deletion of a node was proposed in \cite{3}, the notion of level
was inherent in the first $E_{11}$ paper \cite{1}, but was spelt out explicitly together with the constraints
on the representations in the context of $E_{10}$ in \cite{E10} and in general in \cite{A8}.
\par
At level one, that is $m_c=1$, equation (A.5) becomes  $\sum_i q_i \mu_i - \nu_1=-\sum_i n_i\beta_i$ which
lies in the negative root lattice. One obvious solution is that $\sum_i q_i \mu_i =  \nu_1$. An identical
discussion applies to equation (A.7). Hence at level one we always have the representation of $G_1\otimes
G_2$ with highest weights $(\nu_1,\nu_2)$ in the adjoint representation of $G^{+++}$.
\par
We now specialise to the  cases concerning  the very extended algebra $E_{11}$, or $E_8^{+++}$, whose Dynkin
diagram in given in figure 1. The theory in $D$ dimensions is found by deleting the node labelled $D$ and
decomposing into representations of the remaining algebra which is $A_{D-1} \otimes G_2$. The algebra
$A_{D-1}$, or $SL(D)$, corresponds in the non-linear realisation to the gravity sector. In these cases the
deleted node is attached to the end of the Dynkin diagram of the $A_{D-1}$ subalgebra, that is to the node
labeled $D-1$. As a result we find that $\nu_1 = \mu_{D-1}$. The algebra $G_2$ is in this case $E_{11-D}$ and
the deleted node, $D$ attaches to the first node of this algebra which we label by one. By $E_5$, $E_4$ and
$E_3$ we mean $D_5$,  $A_4$ and $A_2\otimes A_1$ respectively. Once we delete node $D$ we relabel the nodes
of $E_{11-D}$ by $n\to n-D$ to have a sensible labelling from the view point of the subalgebra, as shown in
figure 25. Consequently, we have that $\nu_2 =\lambda_1$. We find that
$$
x^2=1+{1\over D}-\lambda_1^2 \ \ . \eqno(A.10)$$
\par
The level one solution discussed just above is the  representation with highest weight
$(\mu_{D-1},\lambda_1)$. This corresponds to a generator which is a vector under $SL(D)$ and belongs to the
fundamental representation with highest weight $\lambda_1$ under $E_{11-D}$.
\par
Let us first analyse equation (A.6) for the case we are studying here. It becomes
$$\sum_iq_i (\mu_i,\mu_j)-m_c(\mu_{D-1},\mu_j) = -n_j \ \ .
\eqno(A.11)$$ To  analyse this equation it is useful to consider the $SL(D)$ indices that the corresponding
generators carry. The generators are constructed from the multiple commutators of the Chevalley generators.
The generators of $A_{D-1}$ are just $K^a{}_b$ and so the Chevalley generators they contain do not add or
subtract from the total number of indices. However, equation (A.11) is the same equation as one would find if
one analysed the decomposition of the adjoint representations of $A_D$ into $A_{D-1}$ by deleting the end
node {\it i.e.} node $D$. At level $m_c$ the multiple commutator contains $m_c$ Chevalley generators
$K^{D}{}_{D+1}$ associated with the deleted node. As these are   related by the action of $A_{D-1}$ to the
generators $K^{i}{}_{D+1}, i=1,2,\ldots D$ we find that the effect is to lead to a generator that has $m_c$
vector indices. As such, the number of indices on a generator that arises at level $m_c$ is $m_c$. On the
other hand, as the generators are representations of $A_{D-1}$  with Dynkin indices $q_i$ they must have
$\sum_i q_i (D-i)+sD $ indices where the last term corresponds to the possibility of $s$ blocks of $D$
antisymmetrised indices. We recall that having a non-trivial Dynkin index $q_i$ corresponds to having $q_i$
blocks each with $D-i$ totally anti-symmetrised indices. As such we find that
$$ m_c = \sum_i q_i (D-i) + sD \ \ .
\eqno(A.12)$$ It follows that among the solutions that occur to equation (A.11) are all the representations
that occur in the decomposition of the adjoint representation of $A_{D}$ into representations  of $A_{D-1}$.
The actual problem may have more or less solutions as the condition for $\alpha^2$ is different in the latter
case to the problem being studied here.
\par
Substituting the value of $m_c$ of equation (A.12) into equation (A.11) we find that the latter is
automatically solved and that the root coefficients are given by
$$ n_j = s j + \sum_{i \leq j} q_i(j-i) \ \ .
\eqno(A.13)
  $$
The fact that the right-hand side is non-negative indeed implies that the solution always exists. Here we
have used the  formula for the scalar product of fundamental weights of an $A_{D-1}$ algebra
  $$
  (\mu_i,\mu_j) = \begin{cases}{i(D-j) \over D}, \ \ i\leq j\\  {j(D-i) \over
  D}, \ \ j\leq i
  \end{cases} \ \ .
  \eqno(A.14)
  $$

We are interested in forms in terms of their $A_{D-1}$ indices and in particular the representations of
$E_{11-D}$ they belong to.  By a $k$-form we mean an object  that has just one set of  $A_{D-1}$ indices that
is a set of $k$ completely antisymmetrised indices. This is a representation of $SL(D)$ with fundamental
weight $\mu_{D-k}$, which is equivalent to the condition $q_{D-k}=1$, with all other Dynkin indices
vanishing, with the additional requirement that there are no blocks of $D$ indices, {\it i.e.} $s=0$. As such
in equation (A.12)   we require $m_c=k$, $s=0$  and equation (A.11) becomes
  $$(\mu_{D-k},\mu_j)- k (\mu_{D-1},\mu_j) = -n_j
  \eqno(A.15)
  $$
which is automatically solved by taking $s=0$  and $q_{D-k}=1$ in equation (A.13). For the exceptional case
of space filling forms we have one block of $D$ totally anti-symmetrised indices and so $k=D$, $S=1$ and
$m_c=D$, with all the Dynkin indices $q_j$ vanishing.
\par
Rather than solving equation (A.8) for the $E_{11-D}$ highest weights  it is quicker to first solve equation
(A.9). For the groups we are considering and for the case of forms, using equations (A.10) and (A.14) we find
that equation (A.9) becomes for a form or rank $k$, with $k<D$,
  $$
  \alpha^2 = k^2(1-\lambda_1^2) + k + \Lambda^2 \eqno(A.16)
  $$
where $\Lambda^2=\sum_{i,j}p_ip_j(\lambda_i,\lambda_j)$. While for space-filling forms, {\it i.e.} $k=D$ one
has
$$
\alpha^2 = D(D+1) -D^2\lambda_1^2 +  \Lambda^2 \ \ .\eqno(A.17)$$
\par
Since $\alpha^2 =2,0.-2,\ldots$ it is straightforward to find solutions to this equation once we know the
possible scalar products of the fundamental weights which for $E_n, n=6,7,8$ are given by \cite{branes}
  $$((A^{E_n})^{-1})_{ij}=
  \begin{cases}{i(9-n+j)\over (9-n)},\quad
  i,j=1,\ldots,n-3, i\le j\\ {(n-j)((n-3)^2-i(n-5))\over (9-n)},\quad
  i,j=n-3,\ldots,n-1, i\le j \\
  2{i(n-j)\over (9-n)}, \quad i=1,\ldots,n-3, j=n-3,\ldots,n-1, \end{cases}
  \eqno(A.18)$$
and
  $$
  ((A^{E_n})^{-1})_{in}=\begin{cases}{3i\over (9-n)},\quad  i=1,\ldots,n-3\\ {(n-3)(n-i)\over
  (9-n)},\quad
  i=n-3,\ldots,n-1\end{cases} \eqno(A.19)$$ and
  $$
  ((A^{E_n})^{-1})_{nn}= {n\over (9-n)}\ \ . \eqno(A.20)$$
\par
In fact in many cases it suffices to know the length squared of the fundamental weights which are given in
table 1, where the labelling is as in figure 25.
\begin{table}
\begin{center}
\begin{tabular}{|c||c|c|c|c|c|c|c|c|}
\hline \rule[-1mm]{0mm}{6mm}  & $\lambda_1^2$ & $\lambda_2^2$  & $\lambda_3^2$  & $\lambda_4^2$  &
$\lambda_5^2$  & $\lambda_6^2$
& $\lambda_7^2$ & $\lambda_8^2$ \\
\hline \rule[-1mm]{0mm}{6mm} $E_8$ & 2 & 6 & 12 & 20 & 30 & 14 & 4 & 8
 \\
 \cline{1-9} \rule[-1mm]{0mm}{6mm} $E_7$ & 3/2 & 4 & 15/2 & 12 & 6 & 2 & 7/2 \\
 \cline{1-8} \rule[-1mm]{0mm}{6mm} $E_6$  &  4/3 & 10/3 & 6 & 10/3 & 4/3 & 2 \\
\cline{1-7} \rule[-1mm]{0mm}{6mm} $D_5$ & 1 & 2 & 3 & 5/4 & 5/4  \\
 \cline{1-6} \rule[-1mm]{0mm}{6mm} $A_4$ & 4/5 & 6/5 & 6/5 & 4/5 \\
 \cline{1-5}
\end{tabular}
\end{center}
\caption{\small Table giving the square length of the fundamental weights of the internal symmetry groups
occurring in $E_8^{+++}$.\label{Table1}}
\end{table}
To illustrate how this goes let us study the case of $D=5$, that is $E_6$, for which $\alpha^2=-{1\over 3}k^2
+ k + \Lambda^2$ for a $k$-form generator of $A_4$. Taking $k=1$ or $k=2$, we find that $\alpha^2={2\over 3}
+\Lambda^2$ and examining the above table we conclude that $\Lambda^2={4\over 3}$ and that  $p_1=1$, or
$p_5=1$, with all the other Dynkin indices zero. For $k=3$, we find that $\alpha^2=0+\Lambda^2$ and so
$\Lambda^2=2$ or $\Lambda^2=0$ and so $p_6=1$ all the other Dynkin indices zero or we have an $E_6$ singlet.
Finally, for  $k=4$, we find that $\alpha^2=-{4\over 3}+\Lambda^2$ and so $\Lambda^2={4\over 3}$ or
$\Lambda^2={10\over 3}$ and so $p_5=1$ or $p_1=1$ or $p_4=1$ or $p_2=1$ all the other Dynkin indices zero.
\par
We now must check that the above solutions do indeed solve equation (A.8).  For example, for the case of
$D=5$ and so $E_6$, and  taking $k=1$ we find  that each form  belongs to only one fundamental representation
of $E_{11-D}$. The exception is the case of four forms, that is $m_c=1$ for which we have the solutions
$\lambda_1$ and $\lambda_4$. However the former case has an $\alpha^2=0$ and has multiplicity zero.
\par
For the case of space-filling forms for $E_6$ we have $m_c=5$ and $\alpha^2=-{10\over 3}+ \Lambda^2$. The
possible solutions are $\lambda_5$, $\lambda_4$ $\lambda_2$, $\lambda_1$, $2\lambda_4$ and
$\lambda_1+\lambda_6$. Equation (A.8) rules out  $\lambda_4$ and it turns out that $\lambda_1$, $\lambda_2$,
$2\lambda_4$  have multiplicity zero. Hence for the space-filling forms we  find $\lambda_5$ and
$\lambda_1+\lambda_6$.
\par
Let us now consider the case of $D=3$ or $E_8$. In this case from equation (A.16) we have $\alpha^2= - k^2 +k
+\Lambda^2$. Hence for $k=1$ we have $1$ and $\lambda_1$ as usual, while for $k=2$ we can have $1, \lambda_7$
and $\lambda_1$. It turns out the latter and the singlet in $k=1$ have multiplicity zero. For space-filling
branes  we find $\alpha^2=-6+\Lambda^2$ and so we have $1$, $\lambda_1$, $\lambda_2$, $\lambda_7 $,
$\lambda_8$ and $2\lambda_1$. It turns out that $1$ $\lambda_2$ and  $2\lambda_1$ have multiplicity zero.
\par
These results and those for all the other cases are summarised in table 2 \cite{us}.
\begin{table}
\scriptsize
\begin{center}
\begin{tabular}{|c|c||c|c|c|c|c|c|c|}
\hline \rule[-1mm]{0mm}{6mm}
D & G & 1-forms & 2-forms & 3-forms & 4-forms & 5-forms & 6-forms & 7-forms \\
 \hline \rule[-1mm]{0mm}{6mm} \multirow{3}{*}{7} & \multirow{3}{*}{$A_4$} & \multirow{3}{*}{${\bf
{10}}\ (\lambda_1 )$} & \multirow{3}{*}{${\bf \overline{5} }\ (\lambda_3 )$} & \multirow{3}{*}{${\bf {5}} \
(\lambda_4 )$} & \multirow{3}{*}{${\bf \overline{10}} \ (\lambda_2 )$} & \multirow{3}{*}{${\bf 24 } \
(\lambda_3 + \lambda_4 )$} & ${\bf {40}} \ (\lambda_2 + \lambda_3 )$
& ${\bf \overline{70}} \  (2 \lambda_3 + \lambda_4 )$  \\
& & & & & & & & ${\bf  \overline{45}} \ (\lambda_4 + \lambda_2 )$  \\
& & & & & & & ${\bf {15}} \ (2 \lambda_4 ) $ & ${\bf \overline{5}} \ (\lambda_3 )$  \\
 \cline{1-9} \rule[-1mm]{0mm}{6mm}\multirow{3}{*}{6} & \multirow{3}{*}{$D_5$} & \multirow{3}{*}{${\bf
\overline{16}} \ (\lambda_1 )$} & \multirow{3}{*}{${\bf 10 } \ (\lambda_4 )$} & \multirow{3}{*}{${\bf {16} }
\ (\lambda_5 )$} & \multirow{3}{*}{${\bf 45} \ (\lambda_3 ) $} &
\multirow{3}{*}{${\bf \overline{144} } \ (\lambda_4 + \lambda_5 )$} & ${\bf 320} \ (\lambda_3 + \lambda_4 )$  \\
& & & & & & & ${\bf {126}} \ (2 \lambda_5 )$ \\
& & & & & & & ${\bf 10} \ (\lambda_4 ) $ \\
\cline{1-8} \rule[-1mm]{0mm}{6mm} \multirow{2}{*}{5} & \multirow{2}{*}{$E_{6}$} & \multirow{2}{*}{${\bf
\overline{27}} \ (\lambda_1 )$} & \multirow{2}{*}{${\bf {27} } \ (\lambda_5 )$} & \multirow{2}{*}{${\bf 78 }
\ (\lambda_6 )$} & \multirow{2}{*}{${\bf \overline{351}} \ (\lambda_4 )$} &
${\bf {1728}} \ (\lambda_5 + \lambda_6 )$  \\
& & & & & &  ${\bf {27}} \ (\lambda_5 )$  \\
 \cline{1-7} \rule[-1mm]{0mm}{6mm} \multirow{2}{*}{4} & \multirow{2}{*}{$E_{7}$} & \multirow{2}{*}{${\bf
56} \ (\lambda_1 )$} & \multirow{2}{*}{${\bf 133 } \ (\lambda_6 )$} &
\multirow{2}{*}{${\bf 912 }\ (\lambda_7 )$} & ${\bf 8645} \ (\lambda_5 )$ \\
 & & & & & ${\bf 133} \ (\lambda_6 )$ \\
 \cline{1-6} \rule[-1mm]{0mm}{6mm}
 \multirow{3}{*}{3} &  \multirow{3}{*}{$E_{8}$} &  \multirow{3}{*}{${\bf 248} \ (\lambda_1 )$} & ${\bf 3875}
 \ (\lambda_7 ) $
 & ${\bf 147250} \ (\lambda_8)$  \\
 & & & & ${\bf 3875} \ (\lambda_7 )$ \\
 & & & ${\bf 1} \ (0)$ & ${\bf 248} \ (\lambda_1 ) $  \\
 \cline{1-5}
\end{tabular}
\end{center}
\caption{\small Table giving the representations of the symmetry group $G$ of all the generators with
completely antisymmetric indices of $E_8^{+++}$ in dimension from 7 to 3, and the corresponding highest
weight. The representations are the conjugates of the ones in table 5 of \cite{us}, where the corresponding
fields were listed.  \label{Table2}}
\end{table}
Once the results are listed in terms of their highest weights a pattern for all the groups is apparent. The
1-form generators always have highest weight $\lambda_1$. Indeed these generators are the level one
generators, and this is the representation with highest weight $\nu_2$ already discussed. The 2-form
generators in $D$ dimensions belong to the representation of $E_{11-D}$ with highest weight $\lambda_{10-D}$.
The three-dimensional case in exceptional because together with $\lambda_7$, which follows the patters, one
also gets a singlet of $E_8$. The 3-forms always contain the representation with highest weight
$\lambda_{11-D}$, and the 4-forms always contain the representation with highest weight $\lambda_{9-D}$. The
5-forms always contain the representation with highest weight $\lambda_{10-D} + \lambda_{11-D}$, and the
6-forms always contain the representation with highest weight $\lambda_{9-D} + \lambda_{10-D}$ and the one
with highest weight $2 \lambda_{11-D}$. There is also an additional pattern involving the spacetime-filling
forms, that always contain the representation with highest weight $\lambda_{10-D}$. It is amusing to draw the
Dynkin diagrams of $E_{11-D}$ and place the forms against the node corresponding to the highest weight of the
representations to which it belongs.
\par
The reader can apply the above technology to find the representations of the forms in the other cases
required in this paper. For example one can consider the case of $D_{{P\over 2}+4}^{+++}$ of figure 4
discussed in section 4.3, which deleting node 6 leads to $G_1=SO(6,6)$ and $G_2=SO(P)$. As explained above
the lowest level representation has the highest weight $(\mu_5, \lambda_1)$ and the reader will readily find
the higher level results used in this paper. However, in this case one must further decompose $SO(6,6)$ into
$SL(6)$ to find the recognisable representations of the forms in six dimensions.
\par
One can calculate the representations found in this appendix using the programme SimpLie
\cite{ericembedding}. This has the advantage that it gives the multiplicities of each representation.
However, we think it is useful to give a purely algebraic method that can be carried out by hand. By doing
such calculations one can spot features that one might otherwise miss such as the above pattern of highest
weights. These calculations can also be applied to cases where one wants to compute the representations of
the forms arising in the non-linear realisation of groups like $D_{n}^{+++}$ for arbitrary $n$.

\end{appendix}

\newpage

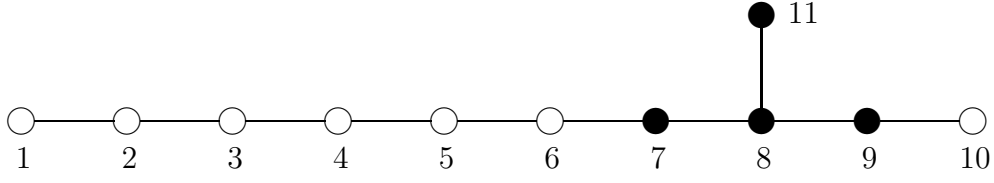
\begin{figure}[h!]
\begin{center}
\begin{picture}(380,60)
\multiput(10,10)(40,0){6}{\circle{10}} \multiput(250,10)(40,0){3}{\circle*{10}} \put(370,10){\circle{10}}
\multiput(15,10)(40,0){9}{\line(1,0){30}} \put(290,50){\circle*{10}} \put(290,15){\line(0,1){30}}
\put(8,-8){$1$} \put(48,-8){$2$} \put(88,-8){$3$} \put(128,-8){$4$} \put(168,-8){$5$} \put(208,-8){$6$}
\put(248,-8){$7$} \put(288,-8){$8$} \put(328,-8){$9$} \put(365,-8){$10$} \put(300,47){$11$}
\end{picture}
\caption{The $E_{8(-24)}^{+++}$ Dynkin diagram corresponding to $L(8,1)$.}
\end{center}
\end{figure}

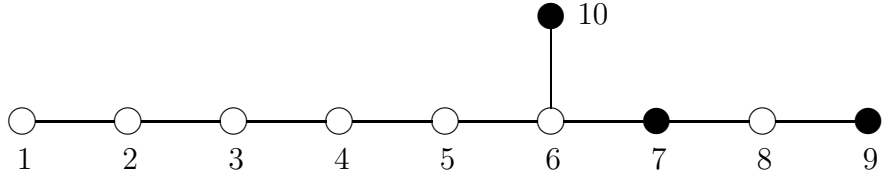
\begin{figure}[h!]
\begin{center}
\begin{picture}(340,60)
  \multiput(10,10)(40,0){6}{\circle{10}}
  \put(250,10){\circle*{10}}
  \put(290,10){\circle{10}}
  \put(330,10){\circle*{10}}
  \put(210,50){\circle*{10}}
  \multiput(15,10)(40,0){8}{\line(1,0){30}}
  \put(210,15){\line(0,1){30}}
  \put(8,-8){$1$} \put(48,-8){$2$} \put(88,-8){$3$} \put(128,-8){$4$} \put(168,-8){$5$} \put(208,-8){$6$}
  \put(248,-8){$7$} \put(288,-8){$8$} \put(328,-8){$9$}
  \put(220,47){$10$}
\end{picture}
\caption{The $E_{7(-5)}^{+++}$ Dynkin diagram corresponding to $L(4,1)$.}
\end{center}
\end{figure}

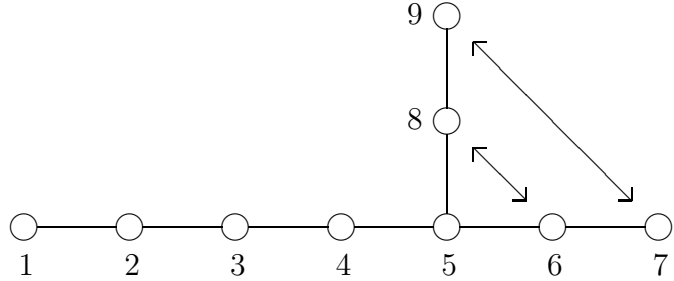
\begin{figure}[h!]
\begin{center}
\begin{picture}(260,100)
  \multiput(10,10)(40,0){7}{\circle{10}}
  \multiput(15,10)(40,0){6}{\line(1,0){30}}
  \multiput(170,50)(0,40){2}{\circle{10}}
%  \multiput(170,130)(0,40){2}{\circle{10}}
  \multiput(170,15)(0,40){2}{\line(0,1){30}}
  \put(8,-8){$1$} \put(48,-8){$2$} \put(88,-8){$3$} \put(128,-8){$4$} \put(168,-8){$5$} \put(208,-8){$6$}
  \put(248,-8){$7$}
  \put(155,47){$8$}
  \put(155,87){$9$}
  \put(200,20){\line(-1,1){20}}
  \put(200,20){\line(-1,0){5}}
  \put(200,20){\line(0,1){5}}
  \put(240,20){\line(-1,0){5}}
  \put(240,20){\line(0,1){5}}
  \put(240,20){\line(-1,1){60}}
  \put(180,40){\line(1,0){5}}
  \put(180,40){\line(0,-1){5}}
  \put(180,80){\line(1,0){5}}
  \put(180,80){\line(0,-1){5}}
 \end{picture}
\end{center}
\caption{The $E_{6(2)}^{+++}$ Dynkin diagram corresponding to $L(2,1)$. } \label{figDL21}
\end{figure}

\begin{figure}[h!]
\begin{center}
\begin{picture}(380,100)
  \multiput(10,50)(40,0){6}{\circle{10}}
  \put(130,90){\circle{10}}
  \put(250,50){\circle*{10}}
  \put(330,50){\circle*{10}}
  \put(370,10){\circle*{10}}
  \put(370,90){\circle*{10}}
  \multiput(15,50)(40,0){6}{\line(1,0){30}}
  \put(130,55){\line(0,1){30}}
  \put(255,50){\line(1,0){20}}
  \put(278,50){\line(1,0){10}}
  \put(290,50){\line(1,0){10}}
  \put(303,50){\line(1,0){22}}
  \put(330,50){\line(1,1){40}}
  \put(330,50){\line(1,-1){40}}
  \put(8,32){$1$} \put(48,32){$2$} \put(88,32){$3$} \put(128,32){$4$} \put(168,32){$5$} \put(208,32){$6$}
  \put(248,32){$7$} \put(308,32){${P \over 2}+4$} \put(368,-8){${P \over 2}+5$} \put(368,72){${P \over 2}+6$}
  \put(140,87){${P \over 2}+7$}
\end{picture}
\end{center}
\caption{The $D^{+++}_{{P \over 2}+4(4)}$  Dynkin diagram corresponding to $L(0,P)$ ($P$ even).}
\end{figure}
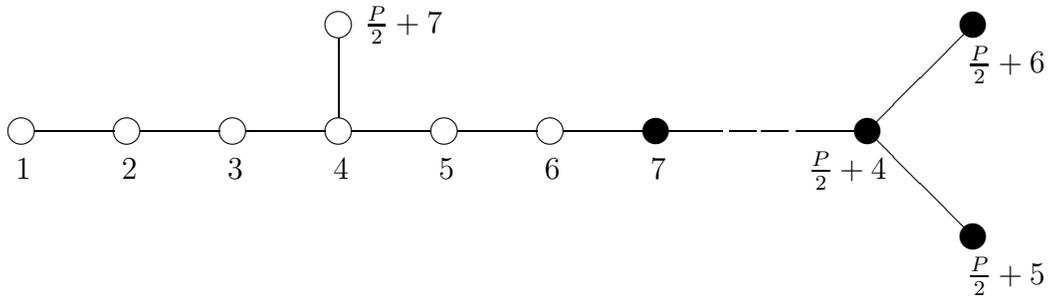

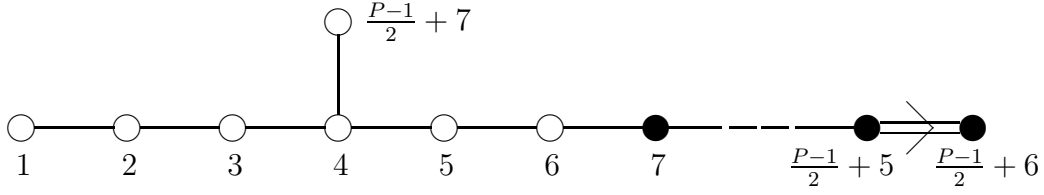
\begin{figure}[h!]
\begin{center}
\begin{picture}(380,60)
  \multiput(10,10)(40,0){6}{\circle{10}}
  \put(130,50){\circle{10}}
  \put(250,10){\circle*{10}}
  \put(330,10){\circle*{10}}
  \put(370,10){\circle*{10}}
  \multiput(15,10)(40,0){6}{\line(1,0){30}}
  \put(130,15){\line(0,1){30}}
  \put(255,10){\line(1,0){20}}
  \put(278,10){\line(1,0){10}}
  \put(290,10){\line(1,0){10}}
  \put(303,10){\line(1,0){22}}
  \put(335,12){\line(1,0){30}}
  \put(335,8){\line(1,0){30}}
  \put(355,10){\line(-1,-1){10}}
  \put(355,10){\line(-1,1){10}}
  \put(8,-8){$1$} \put(48,-8){$2$} \put(88,-8){$3$} \put(128,-8){$4$} \put(168,-8){$5$} \put(208,-8){$6$}
  \put(248,-8){$7$}
  \put(300,-8){${P -1 \over 2} +5$}  \put(355,-8){${P -1 \over 2} +6$} \put(140,47){${P - 1 \over 2} +7$}
\end{picture}
\end{center}
\caption{The $B^{+++}_{{P-1 \over 2}+4(4)}$  Dynkin diagram corresponding to $L(0,P)$ ($P$ odd).}
\end{figure}

\begin{figure}[h!]
\begin{center}
\begin{picture}(380,60)
  \multiput(10,10)(40,0){6}{\circle{10}} \multiput(250,10)(40,0){3}{\circle*{10}} \put(370,10){\circle{10}}
  \multiput(15,10)(40,0){9}{\line(1,0){30}} \put(290,50){\circle*{10}} \put(290,15){\line(0,1){30}}
  \put(205,5){\line(1,1){10}} \put(215,5){\line(-1,1){10}}
  % \put(240,0){\line(1,0){140}}
  % \put(240,0){\line(0,1){60}} \put(240,60){\line(1,0){140}} \put(380,0){\line(0,1){60}}
  \put(8,-8){$1$} \put(48,-8){$2$} \put(88,-8){$3$} \put(128,-8){$4$} \put(168,-8){$5$} \put(208,-8){$6$}
  \put(248,-8){$7$} \put(288,-8){$8$} \put(328,-8){$9$} \put(365,-8){$10$} \put(300,47){$11$}
\end{picture}
\caption{The $E_{8(-24)}^{+++}$ Dynkin diagram corresponding to the 6-dimensional $L(8,1)$ theory. The
internal symmetry group is $SO(9,1)$.}
\end{center}
\end{figure}
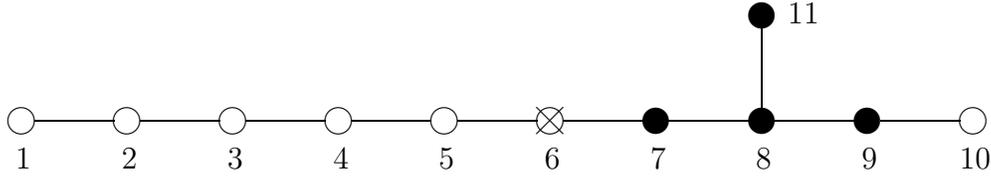

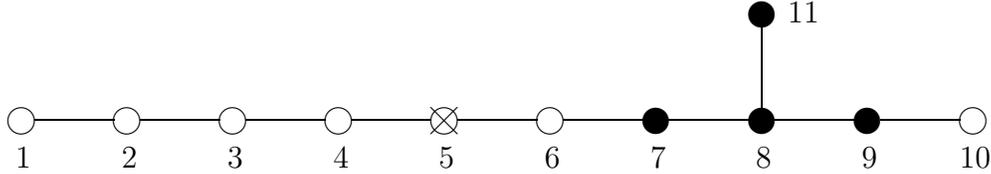
\begin{figure}[h!]
\begin{center}
\begin{picture}(380,60)
  \multiput(10,10)(40,0){6}{\circle{10}} \multiput(250,10)(40,0){3}{\circle*{10}} \put(370,10){\circle{10}}
  \multiput(15,10)(40,0){9}{\line(1,0){30}} \put(290,50){\circle*{10}} \put(290,15){\line(0,1){30}}
  \put(165,5){\line(1,1){10}} \put(175,5){\line(-1,1){10}}
  % \put(200,0){\line(1,0){180}}
  % \put(200,0){\line(0,1){60}} \put(200,60){\line(1,0){180}} \put(380,0){\line(0,1){60}}
  \put(8,-8){$1$} \put(48,-8){$2$} \put(88,-8){$3$} \put(128,-8){$4$} \put(168,-8){$5$} \put(208,-8){$6$}
  \put(248,-8){$7$} \put(288,-8){$8$} \put(328,-8){$9$} \put(365,-8){$10$} \put(300,47){$11$}
\end{picture}
\caption{The $E_{8(-24)}^{+++}$ Dynkin diagram corresponding to the 5-dimensional $L(8,1)$ theory. The
internal symmetry group is $E_{6(-26)}$.}
\end{center}
\end{figure}

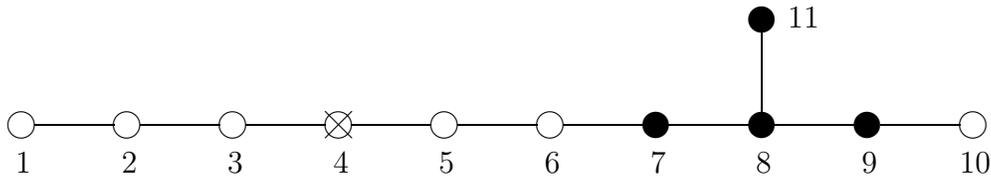
\begin{figure}[h!]
\begin{center}
\begin{picture}(380,60)
  \multiput(10,10)(40,0){6}{\circle{10}} \multiput(250,10)(40,0){3}{\circle*{10}} \put(370,10){\circle{10}}
  \multiput(15,10)(40,0){9}{\line(1,0){30}} \put(290,50){\circle*{10}} \put(290,15){\line(0,1){30}}
  \put(125,5){\line(1,1){10}} \put(135,5){\line(-1,1){10}}
  % \put(160,0){\line(1,0){220}}
  % \put(160,0){\line(0,1){60}} \put(160,60){\line(1,0){220}} \put(380,0){\line(0,1){60}}
  \put(8,-8){$1$} \put(48,-8){$2$} \put(88,-8){$3$} \put(128,-8){$4$} \put(168,-8){$5$} \put(208,-8){$6$}
  \put(248,-8){$7$} \put(288,-8){$8$} \put(328,-8){$9$} \put(365,-8){$10$} \put(300,47){$11$}
\end{picture}
\caption{The $E_{8(-24)}^{+++}$ Dynkin diagram corresponding to the 4-dimensional $L(8,1)$ theory. The
internal symmetry group is $E_{7(-25)}$.}
\end{center}
\end{figure}

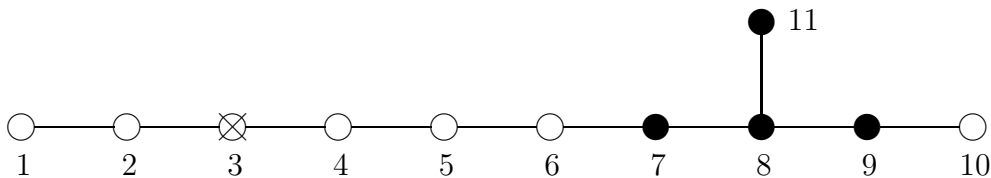
\begin{figure}[h!]
\begin{center}
\begin{picture}(380,60)
  \multiput(10,10)(40,0){6}{\circle{10}} \multiput(250,10)(40,0){3}{\circle*{10}} \put(370,10){\circle{10}}
  \multiput(15,10)(40,0){9}{\line(1,0){30}} \put(290,50){\circle*{10}} \put(290,15){\line(0,1){30}}
  \put(85,5){\line(1,1){10}} \put(95,5){\line(-1,1){10}}
  % \put(120,0){\line(1,0){260}}
  % \put(120,0){\line(0,1){60}} \put(120,60){\line(1,0){260}} \put(380,0){\line(0,1){60}}
  \put(8,-8){$1$} \put(48,-8){$2$} \put(88,-8){$3$} \put(128,-8){$4$} \put(168,-8){$5$} \put(208,-8){$6$}
  \put(248,-8){$7$} \put(288,-8){$8$} \put(328,-8){$9$} \put(365,-8){$10$} \put(300,47){$11$}
\end{picture}
\caption{The $E_{8(-24)}^{+++}$ Dynkin diagram corresponding to the 3-dimensional $L(8,1)$ theory. The
internal symmetry group is $E_{8(-24)}$.}
\end{center}
\end{figure}

\begin{figure}[h!]
\begin{center}
\begin{picture}(340,60)
  \multiput(10,10)(40,0){6}{\circle{10}}
  \put(250,10){\circle*{10}}
  \put(290,10){\circle{10}}
  \put(330,10){\circle*{10}}
  \put(210,50){\circle*{10}}
  \multiput(15,10)(40,0){8}{\line(1,0){30}}
  \put(210,15){\line(0,1){30}}
  \put(205,5){\line(1,1){10}} \put(215,5){\line(-1,1){10}}
  % \put(200,40){\line(1,0){20}}
  % \put(200,60){\line(1,0){20}}
  % \put(200,40){\line(0,1){20}}
  % \put(220,40){\line(0,1){20}}
  % \put(240,0){\line(1,0){100}}
  % \put(240,0){\line(0,1){20}}
  % \put(240,20){\line(1,0){100}}
  % \put(340,0){\line(0,1){20}}
  \put(8,-8){$1$} \put(48,-8){$2$} \put(88,-8){$3$} \put(128,-8){$4$} \put(168,-8){$5$} \put(208,-8){$6$}
  \put(248,-8){$7$} \put(288,-8){$8$} \put(328,-8){$9$}
  \put(220,47){$10$}
\end{picture}
\caption{The $E_{7(-5)}^{+++}$ Dynkin diagram corresponding to the 6-dimensional $L(4,1)$ theory. The
internal symmetry group is $SU(2)\otimes SU^*(4)$.}
\end{center}
\end{figure}
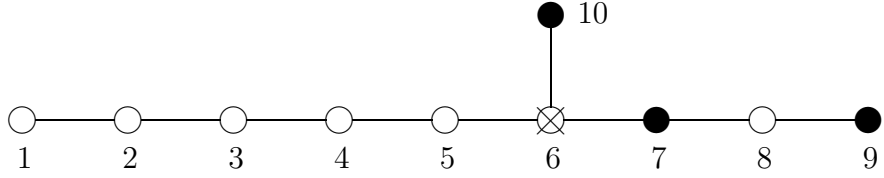

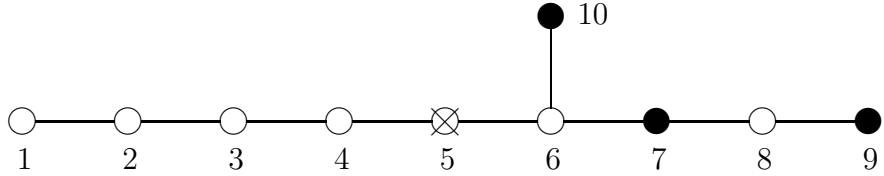
\begin{figure}[h!]
\begin{center}
\begin{picture}(340,60)
  \multiput(10,10)(40,0){6}{\circle{10}}
  \put(250,10){\circle*{10}}
  \put(290,10){\circle{10}}
  \put(330,10){\circle*{10}}
  \put(210,50){\circle*{10}}
  \multiput(15,10)(40,0){8}{\line(1,0){30}}
  \put(210,15){\line(0,1){30}}
  \put(165,5){\line(1,1){10}} \put(175,5){\line(-1,1){10}}
  % \put(200,0){\line(1,0){140}}
  % \put(200,0){\line(0,1){60}}
  % \put(200,60){\line(1,0){140}}
  % \put(340,0){\line(0,1){60}}
  \put(8,-8){$1$} \put(48,-8){$2$} \put(88,-8){$3$} \put(128,-8){$4$} \put(168,-8){$5$} \put(208,-8){$6$}
  \put(248,-8){$7$} \put(288,-8){$8$} \put(328,-8){$9$}
  \put(220,47){$10$}
\end{picture}
\caption{The $E_{7(-5)}^{+++}$ Dynkin diagram corresponding to the 5-dimensional $L(4,1)$ theory. The
internal symmetry group is $SU^*(6)$.}
\end{center}
\end{figure}

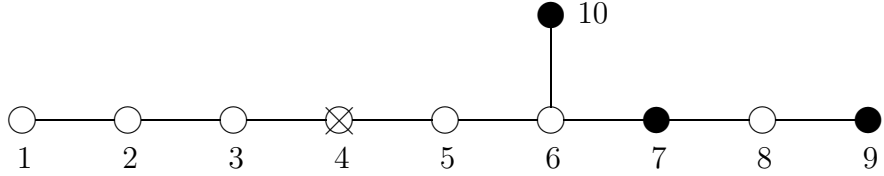
\begin{figure}[h!]
\begin{center}
\begin{picture}(340,60)
  \multiput(10,10)(40,0){6}{\circle{10}}
  \put(250,10){\circle*{10}}
  \put(290,10){\circle{10}}
  \put(330,10){\circle*{10}}
  \put(210,50){\circle*{10}}
  \multiput(15,10)(40,0){8}{\line(1,0){30}}
  \put(210,15){\line(0,1){30}}
  \put(125,5){\line(1,1){10}} \put(135,5){\line(-1,1){10}}
  % \put(160,0){\line(1,0){180}}
  % \put(160,0){\line(0,1){60}}
  % \put(160,60){\line(1,0){180}}
  % \put(340,0){\line(0,1){60}}
  \put(8,-8){$1$} \put(48,-8){$2$} \put(88,-8){$3$} \put(128,-8){$4$} \put(168,-8){$5$} \put(208,-8){$6$}
  \put(248,-8){$7$} \put(288,-8){$8$} \put(328,-8){$9$}
  \put(220,47){$10$}
\end{picture}
\caption{The $E_{7(-5)}^{+++}$ Dynkin diagram corresponding to the 4-dimensional $L(4,1)$ theory. The
internal symmetry group is $SO^*(12)$.}
\end{center}
\end{figure}

\begin{figure}[h!]
\begin{center}
\begin{picture}(340,60)
  \multiput(10,10)(40,0){6}{\circle{10}}
  \put(250,10){\circle*{10}}
  \put(290,10){\circle{10}}
  \put(330,10){\circle*{10}}
  \put(210,50){\circle*{10}}
  \multiput(15,10)(40,0){8}{\line(1,0){30}}
  \put(210,15){\line(0,1){30}}
  \put(85,5){\line(1,1){10}} \put(95,5){\line(-1,1){10}}
  % \put(120,0){\line(1,0){220}}
  % \put(120,0){\line(0,1){60}}
  % \put(120,60){\line(1,0){220}}
  % \put(340,0){\line(0,1){60}}
  \put(8,-8){$1$} \put(48,-8){$2$} \put(88,-8){$3$} \put(128,-8){$4$} \put(168,-8){$5$} \put(208,-8){$6$}
  \put(248,-8){$7$} \put(288,-8){$8$} \put(328,-8){$9$}
  \put(220,47){$10$}
\end{picture}
\caption{The $E_{7(-5)}^{+++}$ Dynkin diagram corresponding to the 3-dimensional $L(4,1)$ theory. The
internal symmetry group is $E_{7(-5)}$.}
\end{center}
\end{figure}
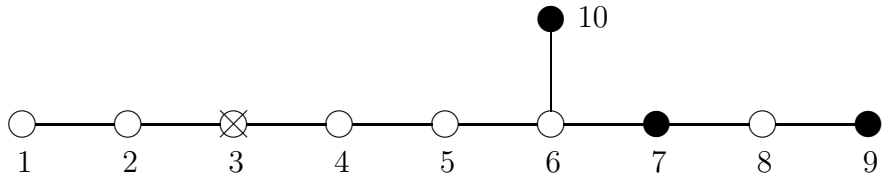

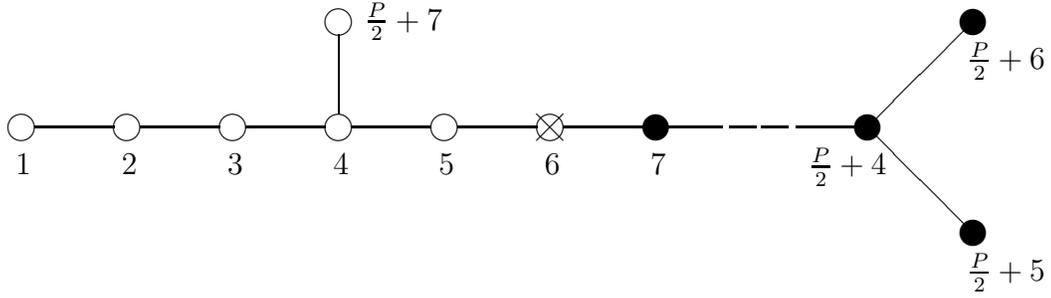
\begin{figure}[h!]
\begin{center}
\begin{picture}(380,100)
  \multiput(10,50)(40,0){6}{\circle{10}}
  \put(130,90){\circle{10}}
  \put(250,50){\circle*{10}}
  \put(330,50){\circle*{10}}
  \put(370,10){\circle*{10}}
  \put(370,90){\circle*{10}}
  \multiput(15,50)(40,0){6}{\line(1,0){30}}
  \put(130,55){\line(0,1){30}}
  \put(255,50){\line(1,0){20}}
  \put(278,50){\line(1,0){10}}
  \put(290,50){\line(1,0){10}}
  \put(303,50){\line(1,0){22}}
  \put(330,50){\line(1,1){40}}
  \put(330,50){\line(1,-1){40}}
%  \put(125,85){\line(1,1){10}}
%  \put(135,85){\line(-1,1){10}}
  \put(205,45){\line(1,1){10}}
  \put(215,45){\line(-1,1){10}}
%  \put(240,0){\line(1,0){140}}
%  \put(240,0){\line(0,1){100}}
%  \put(240,100){\line(1,0){140}}
%  \put(380,0){\line(0,1){100}}
  \put(8,32){$1$} \put(48,32){$2$} \put(88,32){$3$} \put(128,32){$4$} \put(168,32){$5$} \put(208,32){$6$}
  \put(248,32){$7$} \put(308,32){${P \over 2}+4$} \put(368,-8){${P \over 2}+5$} \put(368,72){${P \over 2}+6$}
  \put(140,87){${P \over 2}+7$}
\end{picture}
\end{center}
\caption{The $D^{+++}_{{P \over 2}+4(4)}$ Dynkin diagram corresponding to  the 6-dimensional $L(0,P)$ theory
($P$ even) with $P$ vector multiplets and one tensor multiplet. The non-abelian part of the internal symmetry
group is $SO(P)$, which is compact. The gravity line connects nodes 1, 2, 3, 4 and 5.}
\end{figure}

\begin{figure}[h!]
\begin{center}
\begin{picture}(380,100)
  \multiput(10,50)(40,0){6}{\circle{10}}
  \put(130,90){\circle{10}}
  \put(250,50){\circle*{10}}
  \put(330,50){\circle*{10}}
  \put(370,10){\circle*{10}}
  \put(370,90){\circle*{10}}
  \multiput(15,50)(40,0){6}{\line(1,0){30}}
  \put(130,55){\line(0,1){30}}
  \put(255,50){\line(1,0){20}}
  \put(278,50){\line(1,0){10}}
  \put(290,50){\line(1,0){10}}
  \put(303,50){\line(1,0){22}}
  \put(330,50){\line(1,1){40}}
  \put(330,50){\line(1,-1){40}}
  \put(165,45){\line(1,1){10}}
  \put(175,45){\line(-1,1){10}}
%  \put(200,0){\line(1,0){180}}
%  \put(200,0){\line(0,1){100}}
%  \put(200,100){\line(1,0){180}}
%  \put(380,0){\line(0,1){100}}
  \put(8,32){$1$} \put(48,32){$2$} \put(88,32){$3$} \put(128,32){$4$} \put(168,32){$5$} \put(208,32){$6$}
  \put(248,32){$7$} \put(308,32){${P \over 2}+4$} \put(368,-8){${P \over 2}+5$} \put(368,72){${P \over 2}+6$}
  \put(140,87){${P \over 2}+7$}
\end{picture}
\end{center}
\caption{The $D^{+++}_{{P \over 2}+4(4)}$ Dynkin diagram corresponding to the 6-dimensional $L(0,P)$ theory
($P$ even) with $P+1$ tensor multiplets and no vector multiplets. The internal symmetry group is $SO(P+1,1)$.
The gravity line connects nodes 1, 2, 3, 4 and ${P \over 2}+7$.}
\end{figure}
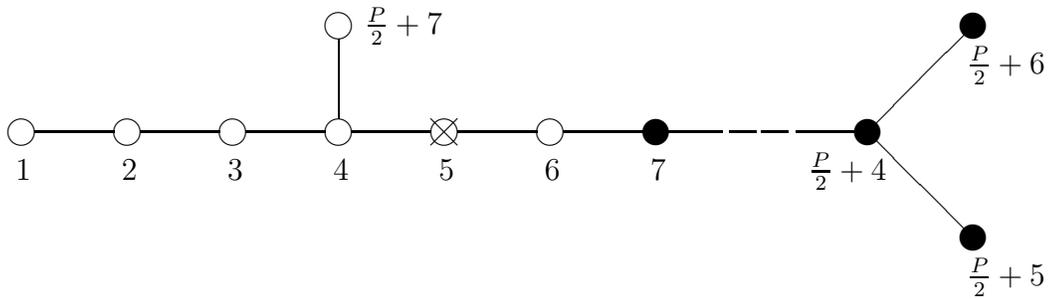

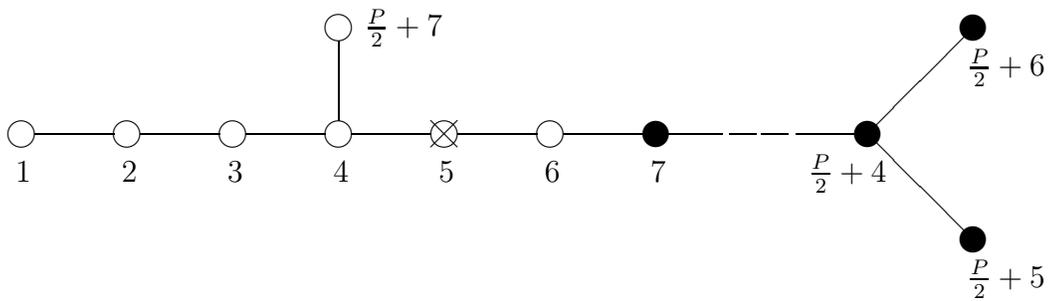
\begin{figure}[h]
\begin{center}
\begin{picture}(380,100)
  \multiput(10,50)(40,0){6}{\circle{10}}
  \put(130,90){\circle{10}}
  \put(250,50){\circle*{10}}
  \put(330,50){\circle*{10}}
  \put(370,10){\circle*{10}}
  \put(370,90){\circle*{10}}
  \multiput(15,50)(40,0){6}{\line(1,0){30}}
  \put(130,55){\line(0,1){30}}
  \put(255,50){\line(1,0){20}}
  \put(278,50){\line(1,0){10}}
  \put(290,50){\line(1,0){10}}
  \put(303,50){\line(1,0){22}}
  \put(330,50){\line(1,1){40}}
  \put(330,50){\line(1,-1){40}}
%  \put(125,85){\line(1,1){10}}
%  \put(135,85){\line(-1,1){10}}
  \put(165,45){\line(1,1){10}}
  \put(175,45){\line(-1,1){10}}
%  \put(200,0){\line(1,0){180}}
%  \put(200,0){\line(0,1){100}}
%  \put(200,100){\line(1,0){180}}
%  \put(380,0){\line(0,1){100}}
  \put(8,32){$1$} \put(48,32){$2$} \put(88,32){$3$} \put(128,32){$4$} \put(168,32){$5$} \put(208,32){$6$}
  \put(248,32){$7$} \put(308,32){${P \over 2}+4$} \put(368,-8){${P \over 2}+5$} \put(368,72){${P \over 2}+6$}
  \put(140,87){${P \over 2}+7$}
\end{picture}
\end{center}
\caption{The $D^{+++}_{{P \over 2}+4(4)}$ Dynkin diagram corresponding to the 5-dimensional $L(0,P)$ theory
($P$ even). The non-abelian part of the internal symmetry group is $SO(P+1,1)$. The gravity line connects
nodes 1, 2, 3 and 4.}
\end{figure}

\begin{figure}[h]
\begin{center}
\begin{picture}(380,100)
  \multiput(10,50)(40,0){6}{\circle{10}}
  \put(130,90){\circle{10}}
  \put(250,50){\circle*{10}}
  \put(330,50){\circle*{10}}
  \put(370,10){\circle*{10}}
  \put(370,90){\circle*{10}}
  \multiput(15,50)(40,0){6}{\line(1,0){30}}
  \put(130,55){\line(0,1){30}}
  \put(255,50){\line(1,0){20}}
  \put(278,50){\line(1,0){10}}
  \put(290,50){\line(1,0){10}}
  \put(303,50){\line(1,0){22}}
  \put(330,50){\line(1,1){40}}
  \put(330,50){\line(1,-1){40}}
  \put(125,45){\line(1,1){10}}
  \put(135,45){\line(-1,1){10}}
%  \put(160,0){\line(1,0){220}}
%  \put(160,0){\line(0,1){100}}
%  \put(160,100){\line(1,0){220}}
%  \put(380,0){\line(0,1){100}}
%  \put(120,80){\line(1,0){20}}
%  \put(120,80){\line(0,1){20}}
%  \put(120,100){\line(1,0){20}}
%  \put(140,80){\line(0,1){20}}
  \put(8,32){$1$} \put(48,32){$2$} \put(88,32){$3$} \put(128,32){$4$} \put(168,32){$5$} \put(208,32){$6$}
  \put(248,32){$7$} \put(308,32){${P \over 2}+4$} \put(368,-8){${P \over 2}+5$} \put(368,72){${P \over 2}+6$}
  \put(140,87){${P \over 2}+7$}
\end{picture}
\end{center}
\caption{The $D^{+++}_{{P \over 2}+4(4)}$ Dynkin diagram corresponding to the 4-dimensional $L(0,P)$ theory
($P$ even). The internal symmetry group is $SO(P+2,2) \otimes SU(1,1)$.}
\end{figure}
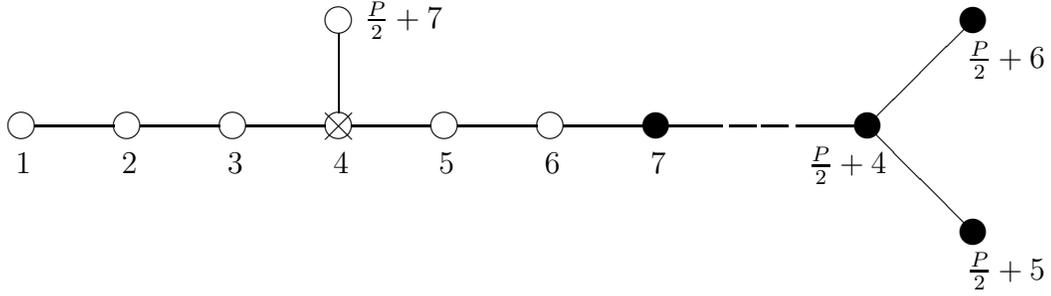

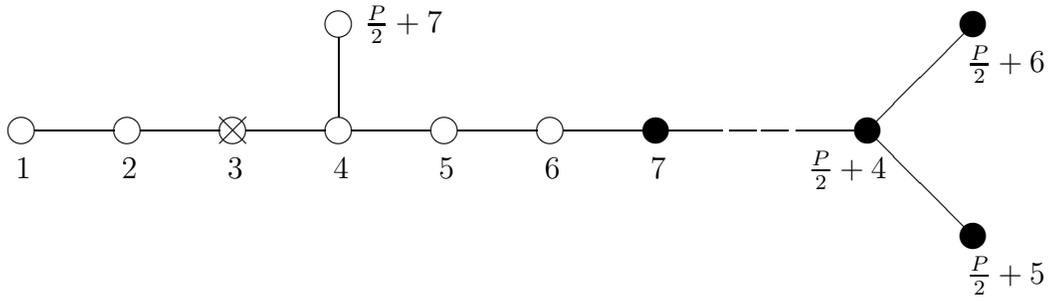
\begin{figure}[h]
\begin{center}
\begin{picture}(380,100)
  \multiput(10,50)(40,0){6}{\circle{10}}
  \put(130,90){\circle{10}}
  \put(250,50){\circle*{10}}
  \put(330,50){\circle*{10}}
  \put(370,10){\circle*{10}}
  \put(370,90){\circle*{10}}
  \multiput(15,50)(40,0){6}{\line(1,0){30}}
  \put(130,55){\line(0,1){30}}
  \put(255,50){\line(1,0){20}}
  \put(278,50){\line(1,0){10}}
  \put(290,50){\line(1,0){10}}
  \put(303,50){\line(1,0){22}}
  \put(330,50){\line(1,1){40}}
  \put(330,50){\line(1,-1){40}}
  \put(85,45){\line(1,1){10}}
  \put(95,45){\line(-1,1){10}}
%  \put(120,0){\line(1,0){260}}
%  \put(120,0){\line(0,1){100}}
%  \put(120,100){\line(1,0){260}}
%  \put(380,0){\line(0,1){100}}
  \put(8,32){$1$} \put(48,32){$2$} \put(88,32){$3$} \put(128,32){$4$} \put(168,32){$5$} \put(208,32){$6$}
  \put(248,32){$7$} \put(308,32){${P \over 2}+4$} \put(368,-8){${P \over 2}+5$} \put(368,72){${P \over 2}+6$}
  \put(140,87){${P \over 2}+7$}
\end{picture}
\end{center}
\caption{ The $D^{+++}_{{P \over 2}+4(4)}$ Dynkin diagram corresponding to the 3-dimensional $L(0,P)$ theory
($P$ even). The internal symmetry group is $SO(P+4,4)$.}
\end{figure}

\newpage

\begin{figure}[h]
\begin{center}
\begin{picture}(420,20)
  \multiput(10,10)(40,0){3}{\circle{10}}
  \put(130,10){\circle*{10}}
  \put(170,10){\circle{10}}
  \multiput(210,10)(40,0){2}{\circle*{10}}
  \multiput(330,10)(40,0){3}{\circle*{10}}
  \multiput(15,10)(40,0){2}{\line(1,0){30}}
  \multiput(135,10)(40,0){3}{\line(1,0){30}}
  \put(95,12){\line(1,0){30}}
  \put(95,8){\line(1,0){30}}
  \put(115,10){\line(-1,-1){10}}
  \put(115,10){\line(-1,1){10}}
  \put(335,10){\line(1,0){30}}
  \put(375,12){\line(1,0){30}}
  \put(375,8){\line(1,0){30}}
  \put(385,10){\line(1,1){10}}
  \put(385,10){\line(1,-1){10}}
  \put(255,10){\line(1,0){20}}
  \put(278,10){\line(1,0){10}}
  \put(291,10){\line(1,0){10}}
  \put(305,10){\line(1,0){20}}
  \put(8,-8){$1$} \put(48,-8){$2$} \put(88,-8){$3$} \put(128,-8){$4$} \put(168,-8){$5$} \put(208,-8){$6$}
  \put(248,-8){$7$}  \put(315,-8){$P+3$} \put(355,-8){$P+4$} \put(395,-8){$P+5$}
\end{picture}
\caption{The $C_{P+2}^{+++}$ Dynkin diagram corresponding to $L(-3,P)$.}
\end{center}
\end{figure}
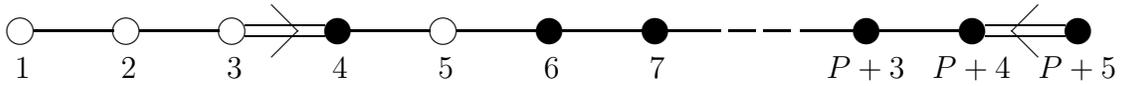

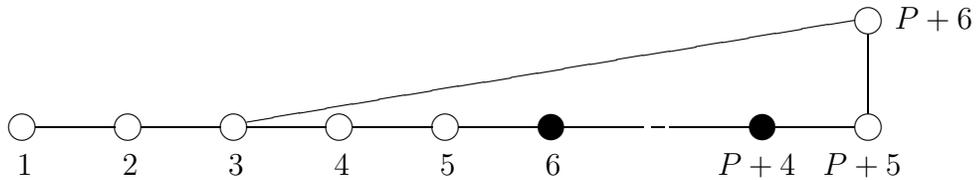
\begin{figure}[h!]
\begin{center}
\begin{picture}(340,60)
  \multiput(10,10)(40,0){5}{\circle{10}}
  \put(210,10){\circle*{10}}
%  \put(250,10){\circle*{10}}
  \put(290,10){\circle*{10}}
  \put(330,10){\circle{10}}
  \put(330,50){\circle{10}}
  \multiput(15,10)(40,0){8}{\line(1,0){30}}
  \put(330,15){\line(0,1){30}}
  \put(95,12){\line(6,1){230}}
  \put(248,10){\line(1,0){5}}
  \put(8,-8){$1$} \put(48,-8){$2$} \put(88,-8){$3$} \put(128,-8){$4$} \put(168,-8){$5$} \put(208,-8){$6$}
  \put(273,-8){$P+4$} \put(313,-8){$P+5$}
  \put(340,47){$P+6$}
\end{picture}
\caption{The $A_{P+3}^{+++}$ Dynkin diagram corresponding to $L(-2,P)$. All nodes from 6 to $P+4$ are
black. Nodes 4 and $P+6$ are connected by arrows, as well as nodes 5 and $P+5$.}
\end{center}
\end{figure}

\clearpage

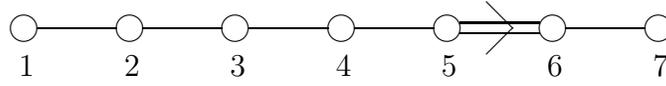
\begin{figure}[h]
\begin{center}
\begin{picture}(260,20)
 \multiput(10,10)(40,0){7}{\circle{10}}
 \multiput(15,10)(40,0){4}{\line(1,0){30}}
% \multiput(170,50)(0,40){2}{\circle{10}}
 \put(175,12){\line(1,0){30}}
 \put(175,8){\line(1,0){30}}
 \put(215,10){\line(1,0){30}}
 \put(195,10){\line(-1,-1){10}}
 \put(195,10){\line(-1,1){10}}
% \put(168,15){\line(0,1){30}}
% \put(172,15){\line(0,1){30}}
% \put(170,55){\line(0,1){30}}
% \put(170,35){\line(1,-1){10}}
% \put(170,35){\line(-1,-1){10}}
   \put(8,-8){$1$} \put(48,-8){$2$} \put(88,-8){$3$} \put(128,-8){$4$} \put(168,-8){$5$} \put(208,-8){$6$}
   \put(248,-8){$7$}
\end{picture}
\end{center}
\caption{The $F_{4(4)}^{+++}$ Dynkin diagram corresponding to $L(1,1)$. }
\end{figure}

\begin{figure}[h]
\begin{center}
\begin{picture}(180,20)
 \multiput(10,10)(40,0){5}{\circle{10}}
 \multiput(15,10)(40,0){4}{\line(1,0){30}}
% \put(130,50){\circle{10}}
% \put(130,15){\line(0,1){30}}
 \put(135,12){\line(1,0){30}}
  \put(135,8){\line(1,0){30}}
  \put(155,10){\line(-1,-1){10}}
  \put(155,10){\line(-1,1){10}}
    \put(8,-8){$1$} \put(48,-8){$2$} \put(88,-8){$3$} \put(128,-8){$4$}
    \put(168,-8){$5$}
\end{picture}
\end{center}
\caption{The $G_{2(2)}^{+++}$ Dynkin diagram.}
\end{figure}
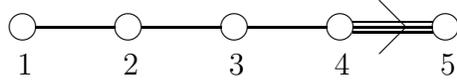

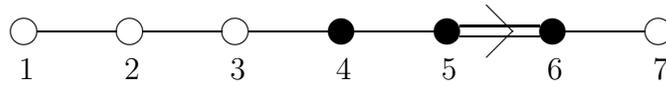
\begin{figure}[h]
\begin{center}
\begin{picture}(260,20)
 \multiput(10,10)(40,0){3}{\circle{10}}
 \multiput(130,10)(40,0){2}{\circle*{10}}
 \multiput(15,10)(40,0){4}{\line(1,0){30}}
 \put(210,10){\circle*{10}}
 \put(250,10){\circle{10}}
 \put(175,12){\line(1,0){30}}
 \put(175,8){\line(1,0){30}}
 \put(215,10){\line(1,0){30}}
 \put(195,10){\line(-1,-1){10}}
 \put(195,10){\line(-1,1){10}}
   \put(8,-8){$1$} \put(48,-8){$2$} \put(88,-8){$3$} \put(128,-8){$4$} \put(168,-8){$5$} \put(208,-8){$6$}
   \put(248,-8){$7$}
\end{picture}
\end{center}
\caption{The $F_{4(-20)}^{+++}$ Dynkin diagram. }
\end{figure}

\begin{figure}[h!]
\begin{center}
\begin{picture}(260,100)
  \multiput(10,10)(40,0){4}{\circle{10}}
  \multiput(170,10)(40,0){2}{\circle*{10}}
  \put(250,10){\circle{10}}
  \multiput(15,10)(40,0){6}{\line(1,0){30}}
  \put(170,50){\circle*{10}}
  \put(170,90){\circle{10}}
%  \multiput(170,130)(0,40){2}{\circle{10}}
  \multiput(170,15)(0,40){2}{\line(0,1){30}}
  \put(8,-8){$1$} \put(48,-8){$2$} \put(88,-8){$3$} \put(128,-8){$4$} \put(168,-8){$5$} \put(208,-8){$6$}
  \put(248,-8){$7$}
  \put(155,47){$8$}
  \put(155,87){$9$}
    \put(240,20){\line(-1,0){5}}
  \put(240,20){\line(0,1){5}}
  \put(240,20){\line(-1,1){60}}
    \put(180,80){\line(1,0){5}}
  \put(180,80){\line(0,-1){5}}
\end{picture}
\end{center}
\caption{The $E_{6(-14)}^{+++}$ Dynkin diagram. }
\end{figure}
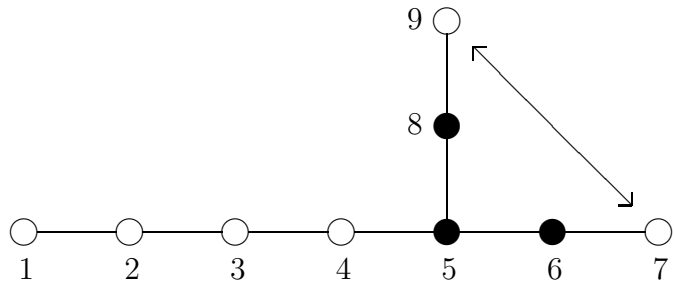
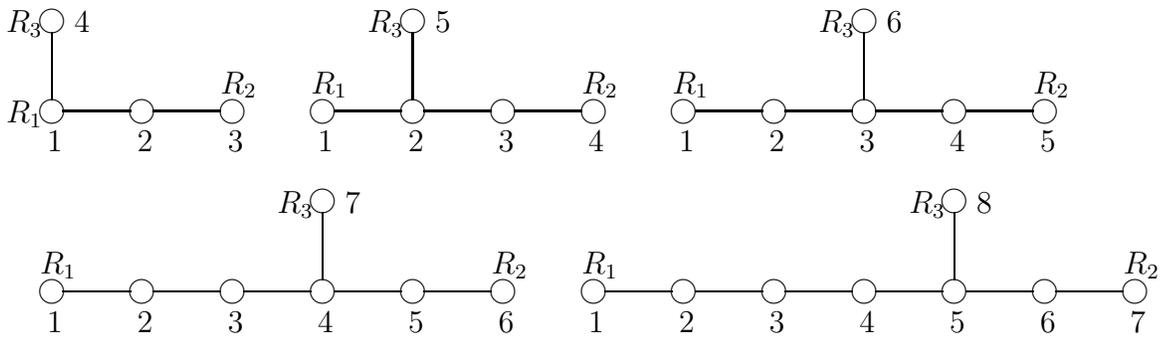
\begin{figure}[h]
\setlength{\unitlength}{0.3mm}
\begin{center}
\begin{picture}(540,140)
  \multiput(10,10)(40,0){13}{\circle{10}}
  \multiput(10,90)(40,0){12}{\circle{10}}
  \put(130,50){\circle{10}}
  \put(410,50){\circle{10}}
  \put(10,130){\circle{10}}
  \put(170,130){\circle{10}}
  \put(370,130){\circle{10}}
  \multiput(15,10)(40,0){5}{\line(1,0){30}}
  \multiput(255,10)(40,0){6}{\line(1,0){30}}
  \multiput(15,90)(40,0){2}{\line(1,0){30}}
  \multiput(135,90)(40,0){3}{\line(1,0){30}}
  \multiput(295,90)(40,0){4}{\line(1,0){30}}
  \put(10,95){\line(0,1){30}}
  \put(130,15){\line(0,1){30}}
  \put(410,15){\line(0,1){30}}
  \put(170,95){\line(0,1){30}}
  \put(370,95){\line(0,1){30}}
  \put(8,-8){$1$} \put(48,-8){$2$} \put(88,-8){$3$} \put(128,-8){$4$} \put(168,-8){$5$} \put(208,-8){$6$}
  \put(248,-8){$1$} \put(288,-8){$2$} \put(328,-8){$3$} \put(368,-8){$4$} \put(408,-8){$5$} \put(448,-8){$6$}
  \put(488,-8){$7$} \put(8,72){$1$} \put(48,72){$2$} \put(88,72){$3$} \put(128,72){$1$} \put(168,72){$2$}
  \put(208,72){$3$} \put(248,72){$4$}\put(288,72){$1$} \put(328,72){$2$} \put(368,72){$3$} \put(408,72){$4$}
  \put(448,72){$5$}
  \put(20,125){$4$} \put(180,125){$5$} \put(380,125){$6$}
  \put(140,45){$7$} \put(420,45){$8$}
  \put(-10,85){$R_1$} \put(85,98){$R_2$} \put(205,18){$R_2$} \put(485,18){$R_2$} \put(5,18){$R_1$}
  \put(245,18){$R_1$} \put(125,98){$R_1$} \put(245,98){$R_2$} \put(285,98){$R_1$} \put(445,98){$R_2$}
  \put(-10,125){$R_3$} \put(150,125){$R_3$} \put(350,125){$R_3$}
  \put(110,45){$R_3$} \put(390,45){$R_3$}
%  \put(248,-8){$7$}
%  \put(155,47){$8$}
%  \put(155,87){$9$}
\end{picture}
\end{center}
\caption{The Dynkin diagrams for $A_4$, $D_5$, $E_6$, $E_7$ and $E_8$ with the labelling of the nodes as in
Appendix A. The pattern for the forms of low rank is particularly apparent when one writes the forms next to
the nodes to which they are associated. }
\end{figure}

\end{document}